\documentclass[acmsmall,screen]{acmart}

\usepackage{url}
\usepackage{xspace}
\usepackage{amsthm}
\usepackage{booktabs}
\usepackage{hyperref}
\usepackage{makecell}
\usepackage{multirow}
\usepackage{graphicx}
\usepackage{enumerate}
\usepackage{todonotes}
\usepackage{color, xcolor}
\usepackage{amsmath, amsfonts}

\newcommand{\ie}{\textit{i.e.,}\xspace}
\newcommand{\eg}{\textit{e.g.,}\xspace}

\newcommand{\etal}{\textit{et al.}\xspace}

\newcommand{\figref}[1]{Fig.~\ref{#1}\xspace}
\newcommand{\tabref}[1]{Table~\ref{#1}\xspace}

\newcommand{\toolname}{{\sc ScenGen}\xspace}
\newcommand{\ob}{\textbf{Observer}\xspace}
\newcommand{\de}{\textbf{Decider}\xspace}
\newcommand{\ex}{\textbf{Executor}\xspace}
\newcommand{\su}{\textbf{Supervisor}\xspace}
\newcommand{\re}{\textbf{Recorder}\xspace}
\newcommand{\cm}{\textit{context memory}\xspace}

\setcopyright{acmlicensed}
\copyrightyear{2026}
\acmYear{2026}
\acmDOI{XXXXXXX.XXXXXXX}

\acmJournal{TOSEM}
\acmVolume{1}
\acmNumber{1}
\acmArticle{1}
\acmMonth{1}

\begin{document}

\title{Scenario-Guided LLM-based Mobile App GUI Testing}

\author{Shengcheng Yu}
\email{shengcheng.yu@tum.de}
\orcid{0000-0003-4640-8637}
\affiliation{\department{School of Computation, Information and Technology, Institute for Advanced Study, Heilbronn Data Science Center, Munich Data Science Institute}\institution{Technical University of Munich}\city{Heilbronn}\country{Germany}}

\author{Yuchen Ling}
\email{yuchenling@smail.nju.edu.cn}
\orcid{0009-0006-9227-3824}
\affiliation{\institution{State Key Laboratory for Novel Software Technology, Nanjing University}\city{Nanjing}\country{China}}

\author{Chunrong Fang}
\authornote{Chunrong Fang is the corresponding author.}
\email{fangchunrong@nju.edu.cn}
\orcid{0000-0003-4640-8637}
\affiliation{\institution{State Key Laboratory for Novel Software Technology, Nanjing University}\city{Nanjing}\country{China}}

\author{Quan Zhou}
\email{qzhou@smail.nju.edu.cn}
\orcid{0000-0000-0000-0000}
\affiliation{\institution{State Key Laboratory for Novel Software Technology, Nanjing University}\city{Nanjing}\country{China}}

\author{Yi Zhao}
\email{zhaoyi@bjev.com.cn}
\orcid{0009-0009-2254-1717}
\affiliation{\institution{State Key Laboratory for Novel Software Technology, Nanjing University}\city{Nanjing}\country{China}}

\author{Chunyang Chen}
\email{chun-yang.chen@tum.de}
\orcid{0000-0003-2011-9618}
\affiliation{\department{School of Computation, Information and Technology, Munich Data Science Institute, Heilbronn Data Science Center, Fortiss}\institution{Technical University of Munich}\city{Heilbronn}\country{Germany}}

\author{Shaomin Zhu}
\email{kerryzhu@tongji.edu.cn}
\orcid{0000-0000-0000-0000}
\affiliation{\institution{Tongji University}\city{Shanghai}\country{Germany}}

\author{Zhenyu Chen}
\email{zychen@nju.edu.cn}
\orcid{0000-0002-9592-7022}
\affiliation{\institution{State Key Laboratory for Novel Software Technology, Nanjing University}\city{Nanjing}\country{China}}

\begin{abstract}

The assurance of mobile application (app) GUI has become increasingly important, as the GUI serves as the primary medium of interaction between users and apps. Although numerous automated GUI testing approaches have been developed with diverse strategies, a substantial gap remains between these approaches and the underlying app business logic. Most existing approaches focus on general exploration rather than the completion of specific testing scenarios, often resulting in missed coverage of critical functionalities. Inspired by the manual testing process, which treats business logic, driven testing scenarios as the fundamental unit of testing, this paper introduces an approach that leverages large language models (LLMs) to comprehend the semantics expressed in app GUIs and their contextual relevance to given testing scenarios. Building upon this capability, we propose \toolname, a novel scenario-guided LLM-based GUI testing framework that employs a multi-agent collaboration mechanism to simulate and automate the phases of manual testing. 

Specifically, \toolname integrates five agents: the \ob, \de, \ex, \su, and \re. The \ob perceives the app GUI state by extracting and structuring GUI widgets and layouts, thereby interpreting the semantic information presented in the GUI. This information is then passed to the \de, which makes scenario-driven decisions with the guidance of LLMs to identify target widgets and determine appropriate actions toward fulfilling specific testing goals. The \ex executes the decided operations on the app, while the \su verifies whether the execution results align with the intended testing scenario completion, ensuring traceability and consistency in test generation and execution. Finally, the \re records the corresponding GUI operations into the \cm as a knowledge base for subsequent decision-making and concurrently monitors runtime bug occurrences. Comprehensive evaluations demonstrate that \toolname effectively generates scenario-guided GUI tests guided by LLM collaboration, achieving higher relevance to app business logic and improving the completeness of automated GUI testing. 

\end{abstract}




\maketitle

\section{Introduction}

With the continuous advancement of mobile devices and mobile Internet technologies, mobile applications (apps) have experienced rapid and large-scale development \cite{yu2024effective}. The accelerated iteration cycles of modern app ecosystems continuously drive their functional and interface evolution. This phenomenon not only enriches user experiences and diversifies app types but also imposes new challenges and higher demands on mobile app quality assurance \cite{yu2024practical}. Among various quality assurance techniques, GUI testing has become indispensable, as the GUI represents the primary interaction layer between users and apps \cite{white2019improving}.

Manual GUI testing, which relies on human testers to explore and interact with the app GUI, remains an intuitive approach for identifying usability issues, interaction anomalies, and visual inconsistencies. However, this process is inherently labor-intensive and time-consuming, making it impractical for frequent app updates and continuous delivery testing scenarios \cite{su2017guided}. Consequently, researchers and practitioners have increasingly turned to automated GUI testing to improve efficiency, scalability, and repeatability. A common paradigm in automated GUI testing is test script record and replay \cite{talebipour2021ui, behrang2019test}, where test scripts are recorded from manual interactions and later executed automatically using frameworks such as Appium \cite{appium}. While this approach enhances execution efficiency, it still heavily depends on manual scripting and frequent maintenance. The tight coupling between scripts and specific GUI structures makes these tests brittle when the app evolves \cite{yoon2022repairing}, leading to high maintenance costs and poor adaptability to new versions or requirements.

To overcome these limitations, automated test generation techniques have emerged. These approaches automatically explore the app under test to generate test cases according to predefined exploration strategies, thereby reducing manual intervention and enhancing testing efficiency. Common strategies include random-based, model-based, and learning-based approaches. Random-based strategies \cite{monkey, machiry2013dynodroid} rely on pseudo-random event generation to explore diverse GUI states, effectively uncovering crashes and inconsistencies. However, they often lack direction and testing scenario awareness. Model-based strategies \cite{su2017guided, mao2016sapienz} improve exploration precision by constructing GUI models through static or dynamic analysis, which are then used to guide event generation. Yet, such models may struggle to capture semantic intent or user-level goals. Learning-based strategies \cite{yu2024effective, pan2020reinforcement, adamo2018reinforcement} leverage advanced AI techniques, such as reinforcement learning, to navigate large and complex GUI state spaces. Although these approaches achieve higher adaptability, they still primarily optimize quantitative coverage metrics rather than aligning with app business logic.

Most existing automated GUI testing approaches are designed to maximize code coverage \cite{yu2024practical}, which, while useful, introduces several limitations. First, generating continuous, logically coherent event sequences that correspond to realistic user behaviors remains difficult \cite{yu2024effective}. As a result, these approaches fail to reproduce complex task flows that represent actual usage patterns. Second, automatically generating contextually appropriate inputs to trigger targeted app responses is still a major challenge \cite{liu2024testing}, leading to superficial interaction exploration that neglects semantic completeness. Third, using code coverage as the primary evaluation metric inadequately reflects the true quality of testing, since it measures execution breadth but not the functional or experiential depth of user interactions \cite{yu2024practical}. The root cause of these limitations lies in the lack of scenario-level understanding of app functionalities \cite{yu2024practical, yu2023llm}. Existing automated approaches focus on state transitions and event diversity rather than on testing scenarios that mirror real user intentions and business workflows. Without such testing scenario comprehension, testing processes are prone to overlooking essential functionalities, resulting in incomplete validation and missed defects \cite{yu2025vision}.

To advance automated GUI testing, there is a pressing need for scenario-aware approaches that can bridge the gap between low-level GUI exploration and high-level business logic. Existing techniques largely operate at the event level and lack the ability to understand the semantics embedded in GUI widgets and layouts, reason about user intentions, and generate coherent test sequences aligned with realistic testing scenarios. Consequently, automated GUI testing must evolve toward LLM-based, human-like reasoning in which each testing scenario is treated as a scenario-guided task consistent with how human testers conceptualize and execute test cases.

\begin{figure}[!htbp]
\centering
\includegraphics[width=\linewidth]{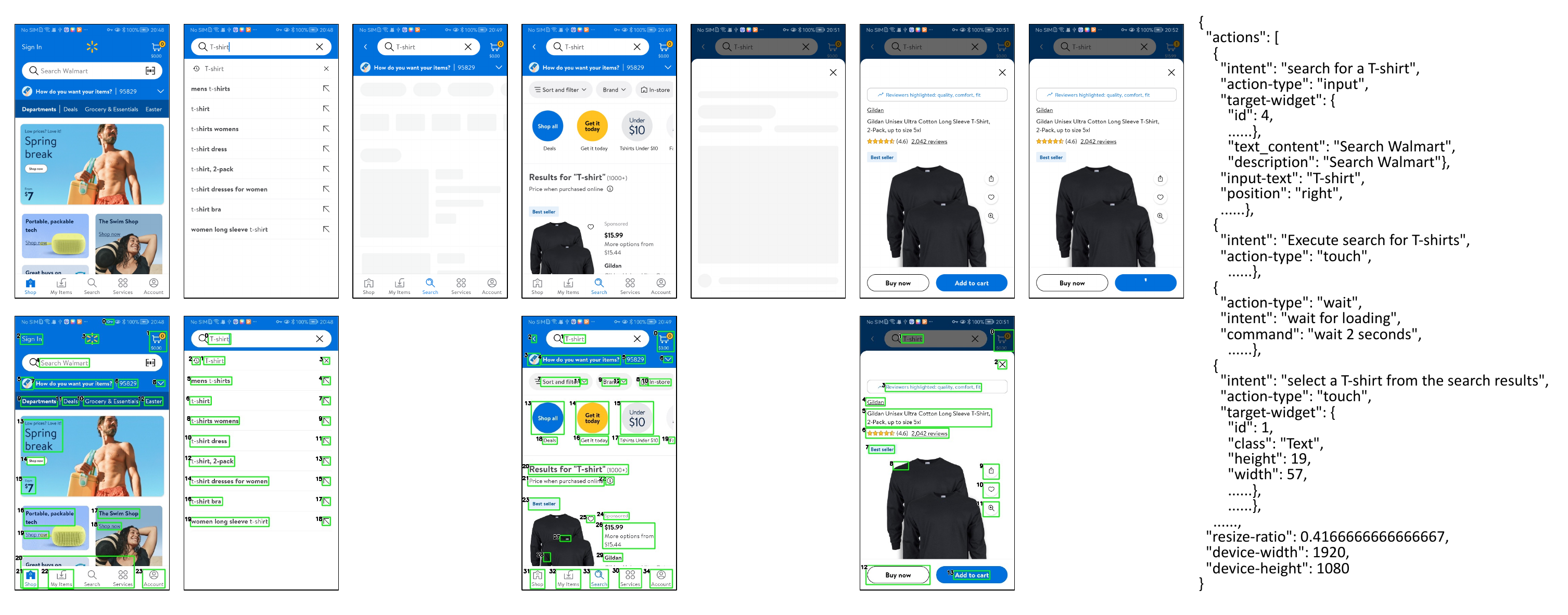}
\caption{Test Generation Example}
\label{fig:example}
\end{figure}

To address these challenges, this paper introduces \toolname, a novel scenario-guided automated GUI testing framework driven by large language models. \toolname integrates LLM-based reasoning into GUI testing to enable semantically meaningful, context-aware, and traceable test generation. Specifically, it bridges GUI-level exploration and scenario-level reasoning by equipping LLMs with the capability to interpret visual semantics of application interfaces and comprehend the contextual intent of testing scenarios. To realize this objective, \toolname employs multi-modal LLMs for semantic perception and adopts a multi-agent collaborative architecture to support intelligent, scenario-consistent test generation. \figref{fig:example} illustrates how \toolname translates a high-level testing scenario, such as completing a shopping task, into a coherent sequence of executable GUI actions. The example also shows how each action is associated with specific widgets identified by the \ob and selected by the \de, forming a structured and traceable action sequence. This example records a shopping testing scenario within the Walmart mobile app. The overall process proceeds as follows: first, the user enters the keyword ``T-shirt'' into the ``Search Walmart'' search bar, then taps a related suggestion or result to execute the search; after waiting for the page to load, the user selects a T-shirt item from the results to open its product detail page; since the first selection may not meet expectations, a back action is performed to return to the search results page, where another T-shirt item is selected; after the page loads again, the user taps the ``Add to cart'' button on the product detail page to add the item to the shopping cart. Each step specifies the action type (input, touch, wait, back), the target GUI element's attributes (such as position, size, and text content), the corresponding ADB command executed, and the associated screenshot path.

The design of \toolname is inspired by the classic OODA Loop decision-making model: Observe, Orient, Decide, and Act, originally proposed for dynamic decision processes in complex environments \cite{boyd1987destruction}. This model aligns well with the iterative nature of GUI testing, where perception, reasoning, execution, and feedback occur cyclically. Drawing from this intuition, \toolname conceptualizes automated GUI testing as an iterative human-like process consisting of five stages: observation, decision-making, execution, result inspection and feedback, and recording. Each stage corresponds to a specific cognitive or operational phase in manual testing, and together they form a closed decision loop that continuously refines the testing process. To operationalize this design, \toolname employs a set of specialized agents, \ob, \de, \ex, \su, and \re, supported by a shared \cm that maintains the overall testing context. Each agent encapsulates distinct reasoning and functional capabilities, enabling modularity, interpretability, and traceability throughout the testing process.

The \ob acts as the perceptual component of \toolname. It identifies and analyzes GUI widgets and layouts, addressing the inherent diversity and complexity of mobile GUIs. Accurate GUI recognition is fundamental for formulating effective testing actions. By combining computer vision and multi-modal LLM techniques, the \ob not only detects GUI components and their attributes but also interprets their semantic roles within the interface (\eg distinguishing between navigation, input, or functional widgets). This design ensures that subsequent decision-making is grounded in a comprehensive and semantically rich understanding of the current app state.

Building upon the output of \ob, the \de functions as the reasoning and planning agent. It interprets the recognized GUI elements in the context of the target testing scenario and generates a detailed action plan for test execution. The design of \de enables independent decision-making by leveraging the testing scenario's semantic goal, rather than random or purely structural exploration, to determine which widgets to interact with and how. This approach provides a logical foundation for scenario-guided testing, ensuring that generated operations align with business logic and the intended testing scenario objectives. Moreover, by decoupling decision-making from perception and execution, \de enhances both the flexibility and generality of \toolname, allowing it to adapt to diverse apps and testing scenarios.

The \ex is responsible for executing the planned operations on the app. By strictly following the action plan generated by \de, the \ex ensures consistency, traceability, and repeatability in testing. Separating execution from decision-making also prevents cascading errors caused by incomplete reasoning or misinterpretation during planning. This modular separation mirrors human testing behavior, where decisions are reasoned out before actual interaction, thereby enhancing testing reliability.

The \su serves as the feedback and validation agent. It inspects whether the executed operations conform to the intended plan and verifies whether the intermediate or final app states satisfy the requirements of the target testing scenario. These requirements are derived from two sources: explicit testing scenario descriptions provided at the start of exploration and implicit semantic understanding embedded within the LLMs. By formalizing this verification process, \su allows \toolname to identify deviations, provide corrective feedback, and maintain a consistent alignment between generated tests and testing scenario objectives. This continuous inspection and correction mechanism prevents error propagation and ensures that the generated tests remain semantically faithful to the intended user flows.

The \re is responsible for recording the complete interaction history, including GUI states, executed actions, and intermediate feedback, into the \cm, which serves as a persistent context memory shared among agents. This design choice enables iterative learning and context-aware decision-making across multiple testing cycles. The recorded context allows \toolname to trace the reasoning behind past actions, reuse relevant knowledge, and dynamically adjust future exploration strategies. Additionally, \re monitors runtime logs and execution traces to identify potential failures or crashes, ensuring that discovered bugs are systematically documented. Once the recording is complete, the next iteration begins until the target testing scenario has been fully explored and validated.

Through this multi-agent, scenario-guided collaboration, \toolname simulates the human tester's reasoning process, observing, interpreting, deciding, acting, and learning iteratively, to achieve semantically coherent, scenario-oriented GUI testing. Prior scenario-based methods typically rely on predefined testing scenario graphs, rule-based transitions, or knowledge graph construction to guide exploration. These approaches depend on static representations of application behavior and are limited by the coverage and completeness of preconstructed models. In contrast, \toolname adopts a multi-agent, scenario-guided architecture that performs dynamic semantic mapping between GUI elements and testing scenario objectives at runtime. Rather than relying on predefined structural models, \toolname interprets GUI layouts in real time through multi-modal reasoning, maintains persistent context memory, and employs a \su-driven feedback loop to ensure testing scenario alignment. This design enables adaptive reasoning under unseen interface variations and dynamic app states. The core difference lies in the shift from model-constrained testing scenario traversal to agentic, semantic-driven exploration coordinated through memory-aware collaboration.

\toolname is designed to address the challenge of scenario-aligned test generation, namely, how to automatically transform a high-level testing scenario into a sequence of low-level GUI interactions that remains consistent with the intended user workflow and the app's business logic. To achieve this, the framework continuously interprets the current GUI state in a structured and semantic manner, reasons about the next action under the guidance of the target scenario, executes the action on the app, and verifies whether the resulting state still advances the scenario as expected. The interaction history and intermediate states are maintained as a testing context so that subsequent decisions remain grounded in prior progress rather than being made in isolation. This design strengthens scenario alignment because each action is generated not only from the current interface state but also from the target scenario and the observed progression of execution, reducing the risk of drifting into irrelevant exploration or stopping at superficially similar but incorrect states. By keeping test generation focused on realistic, business-critical user flows, \toolname produces more coherent and semantically meaningful tests and increases the likelihood of exposing bugs, since many important failures can only be triggered when the app is exercised through the key intermediate steps and deeper functional states required by these scenarios.

In summary, the main contribution of this work is not merely the adoption of computer vision and multi-modal LLMs for GUI understanding, but the design of a testing-oriented, agentic framework that systematically bridges high-level scenario intent and low-level executable GUI actions. \toolname introduces a structured multi-agent architecture in which semantic decision-making is explicitly separated from precise widget localization, and further strengthened through widget prediction and location adjustment mechanisms tailored for automated testing. These mechanisms address the practical gap between semantic intent and pixel-level interaction by refining ambiguous or incomplete visual detections into reliable, executable targets. By integrating scenario-driven reasoning, localization refinement, iterative verification, and structured memory management into a coherent loop, \toolname establishes a principled approach for transforming user-oriented testing scenarios into reproducible and robust GUI test artifacts, thereby advancing scenario-based automation in mobile app testing.

The main contributions of this paper can be summarized as follows:

\begin{itemize}

\item We propose \toolname, a novel scenario-guided, LLM-based GUI testing framework that addresses the challenge of aligning low-level GUI interactions with high-level testing intent and app business logic. The framework integrates semantic GUI understanding, scenario-aware decision making, execution, and verification to generate executable tests that follow predefined testing scenarios rather than unconstrained event exploration.
\item We design a testing-oriented widget localization refinement mechanism that bridges semantic action intent and executable GUI operations. By incorporating widget prediction and location adjustment, the mechanism improves the robustness and precision of scenario-guided test generation under imperfect visual perception.
\item We construct an app benchmark with ten representative testing scenarios and conduct a comprehensive empirical evaluation of \toolname. The results show that \toolname effectively generates scenario-guided GUI tests and improves bug exposure along business-critical user flows.
\end{itemize}

The replication package is available at \textbf{\url{https://github.com/iGUITest/ScenGen}}, and additional implementation details of \toolname are provided at \textbf{\url{https://sites.google.com/view/scengen}}.

\section{Background \& Motivation}

In this section, we first introduce the core concept underlying our work, the testing scenario, which serves as the foundation of the proposed approach. Then, we discuss the background and motivation for employing LLMs in the context of GUI testing. 

\subsection{Testing Scenario}

A testing scenario represents a specific functionality or task within the app that encapsulates a complete and self-consistent business logic from the perspective of end users. Conceptually, a testing scenario reflects how users interact with the app through a coherent sequence of GUI operations to achieve a concrete goal, such as logging in, completing a purchase, or submitting a form. Each testing scenario corresponds to a meaningful user workflow and can be viewed as a semantic unit of testing that connects interface-level actions with business-level objectives.

In real-world app usage, end users interact with mobile apps through a series of GUI operations, such as tapping, swiping, or text input, to accomplish a specific goal. These operation sequences form logical patterns that can be abstracted into corresponding testing scenarios. For example, the ``user login'' testing scenario may include operations like entering credentials, tapping the login button, and verifying access to the home page. Understanding such testing scenarios is critical for testing app functionality from a user-centric perspective, as it enables testers to validate whether the app behaves as intended under realistic usage contexts. In the practice of manual GUI testing, developers and testers typically organize test cases around distinct testing scenarios. This organization allows them to design test cases more systematically, ensuring that all major functionalities and user flows are adequately exercised. Testing scenarios thus serve as a bridge between app functionality and user behavior, facilitating an intuitive and comprehensive verification of the app behavior.

However, despite its practical value, the concept of testing scenarios is largely neglected in existing automated GUI testing approaches. Current methods often emphasize event-level exploration, code coverage, or state-space traversal without explicitly modeling the app's business logic or user goals \cite{yu2024practical}. As a result, these approaches frequently generate fragmented or redundant test sequences that lack semantic coherence. To bridge this gap, there is a need for testing methodologies that can explicitly model and reason about testing scenarios, recognizing not only how to interact with GUI elements but also why those interactions matter in achieving specific business objectives. This observation provides the conceptual motivation for our proposed framework, \toolname, which introduces scenario-guided, LLM-based GUI testing to integrate high-level testing scenario understanding with automated test generation.

\subsection{Motivating Examples}

In this section, we present several real-world examples that illustrate the challenges faced by existing automated GUI testing techniques and motivate the design of \toolname. These examples demonstrate why incorporating the semantic reasoning capabilities of LLMs and grounding test generation in testing scenarios is essential for improving testing completeness and relevance.

\begin{figure}[!htbp]
\centering
\includegraphics[width=\linewidth]{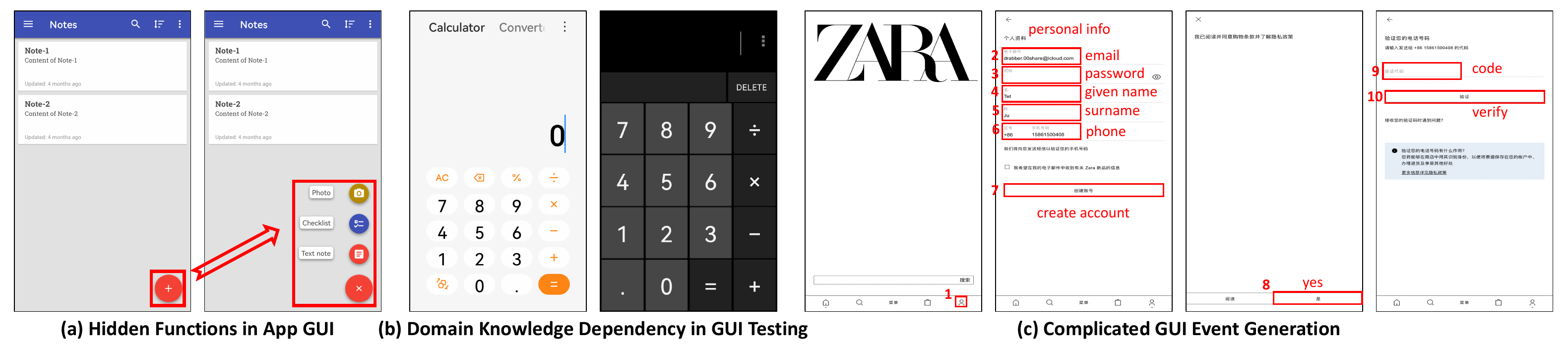}
\caption{Motivating Examples}
\label{fig:mv}
\end{figure}

\subsubsection{Hidden Function Triggering}

Modern app GUIs often employ minimalist design principles, grouping related functionalities behind a single entry to maintain interface simplicity. These hidden widgets are typically accessed through expandable menus or contextual icons. As shown in \figref{fig:mv}(a), in a note-taking app, the widget for creating a ``text note'' is nested within a hidden menu that appears only after clicking the plus icon in the bottom-right corner. If the user clicks outside the menu, the panel automatically closes. Such designs pose a significant challenge for automated GUI testing approaches that rely on surface-level widget exploration or static state transitions, as they may fail to discover the hidden functional paths. During the exploration of the ``create a new note'' testing scenario, LLMs can reason about the GUI's semantic layout and infer that the plus icon is likely associated with content creation. This enables the \ob and \de agents in \toolname to proactively explore the hidden menu and trigger the corresponding functionality, effectively mimicking human intuition in navigating dynamic and context-dependent interfaces.

\subsubsection{Domain Knowledge Dependency}

In many cases, effective GUI testing depends not only on interface exploration but also on domain-specific knowledge relevant to the app's functionality. As illustrated in \figref{fig:mv}(b), although calculator apps have simple GUIs, the correctness of their functionality depends on mathematical boundary conditions and logical rules (\eg division by zero, large number handling). Conventional automated testing methods struggle to identify and generate meaningful test inputs that capture such domain constraints, often limiting their ability to validate correctness comprehensively. By contrast, LLMs possess the capacity to combine GUI understanding with domain reasoning. Within \toolname, the LLM-driven agents can interpret the purpose of calculator widgets, associate them with mathematical operations, and generate scenario-guided test sequences that cover critical boundary and functional conditions. This integration of domain knowledge allows the testing process to move beyond surface-level interactions and focus on semantic correctness, achieving deeper functional coverage.

\subsubsection{Complicated GUI Event Generation}

Many app functionalities are realized through multi-step interaction sequences that reflect the underlying business logic. For instance, as depicted in \figref{fig:mv}(c), a typical login process involves sequential operations such as entering credentials, selecting options, and submitting the form. Each step is temporally dependent and semantically constrained. For example, the ``submit'' button should only be triggered after valid input fields are filled. Traditional automated GUI testing approaches that rely on random or model-based exploration often fail to preserve such logical dependencies, producing invalid or incomplete event sequences. In contrast, \toolname leverages LLMs to reason about the intended workflow and maintain contextual continuity across operations. By understanding the purpose of each widget and its relation to the overall testing scenario goal, LLM agents can generate coherent and meaningful event sequences, such as inputting valid user data followed by appropriate form submission, thus aligning the testing process with realistic user behavior.

\section{Methodology}

\begin{figure}[!htbp]
\centering
\includegraphics[width=\linewidth]{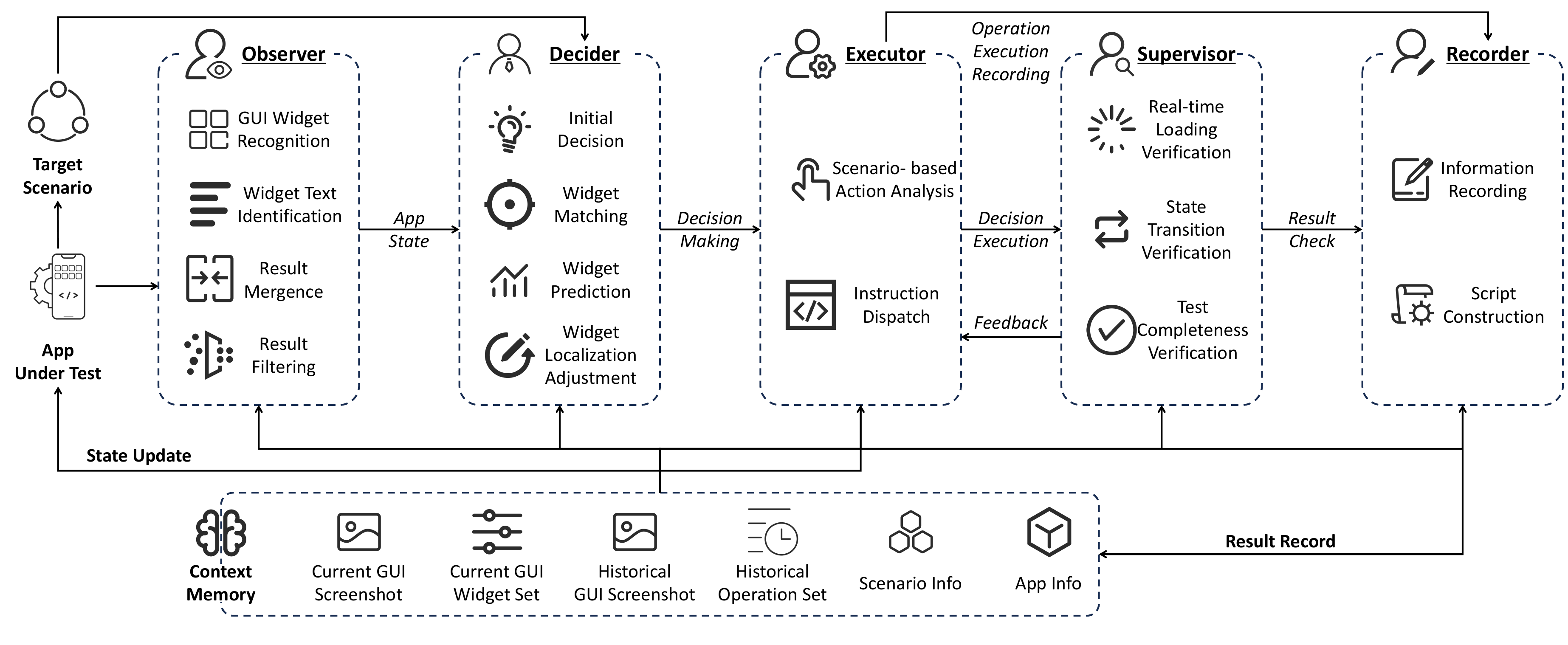}
\caption{\toolname Framework}
\label{fig:framework}
\end{figure}

\toolname comprises multiple collaborative components, as illustrated in \figref{fig:framework}, including the \cm, \ob, \de, \ex, \su, and \re. These components jointly form an iterative process that integrates perception, reasoning, execution, verification, and feedback to achieve scenario-based GUI testing. Among them, several components, such as the \ob, \de, and \su, are powered by the LLMs to provide high-level semantic understanding and decision reasoning to ensure reliable and traceable test execution and validation.

The general input to \toolname is a simple high-level testing scenario description such as ``send an email to a friend'', but directly feeding this description to a single LLM at each interaction step is insufficient for robust GUI testing. A high-level testing scenario captures the testing objective but does not provide structured control over iterative decision-making, state tracking, or execution feedback. Mobile apps often involve dynamic GUI transitions, hidden widgets, asynchronous loading, and unexpected states. Without an explicit mechanism for memory management, structured perception, and verification, repeatedly prompting an LLM with only the testing scenario description would lead to unstable behavior, redundant exploration, or loss of progress context. \toolname addresses this by decomposing the testing process into coordinated agents with persistent context memory, enabling controlled, stepwise reasoning rather than isolated prompt-based inference.

The purpose of generating executable test sequences is to transform abstract testing scenario intent into reproducible, traceable GUI action sequences. A high-level testing scenario alone is not directly executable, nor does it provide visibility into intermediate decisions, widget selections, or state transitions. The generated tests produced by \toolname contain explicit action traces, including widget identifiers, input values, and operation order. This structured output enables reproducibility, debugging, regression testing, and performance comparison across frameworks. Furthermore, the generated sequences allow scenario-level verification through the \su, ensuring that deviations are detected and corrected during execution. In contrast, simply querying an LLM with a testing scenario description lacks traceability, controllability, and systematic validation. The key advantage of test generation lies in converting semantic goals into operational artifacts that can be executed, inspected, and reused.

Specifically, the \ob perceives and interprets the app GUI by detecting widgets and extracting their semantic attributes, enabling the system to understand the current interface context. Building upon this perception, the \de performs reasoning and planning, leveraging LLM capabilities to align GUI interactions with the target testing scenario and business logic. The \ex then executes the planned operations on the app, ensuring repeatability and traceability while decoupling action execution from decision-making to reduce the risk of cascading failures. Next, the \su verifies whether the executed operations achieve the intended testing scenario objectives and provides feedback for correction when discrepancies occur. Finally, the \re records all operations, GUI states, and inspection outcomes into the shared \cm, preserving contextual data to support consistent and informed decision-making in subsequent iterations.

Through the collaboration of these LLM-driven and rule-based components (\ie threshold-based widget mergence, noise filtering, and coordinate extraction), \toolname progressively refines its understanding of app functionality and testing intents until all target testing scenarios are comprehensively explored and validated. To ensure reproducibility and transparency, \textbf{all prompt templates used for agent interactions are publicly available in our online resources.}

\subsection{Context Memory}

The \cm is designed to maintain continuity and coherence throughout the iterative testing process, ensuring that \toolname can generate logically consistent scenario-guided GUI tests. However, large language models (LLMs) are inherently stateless, lacking the ability to retain contextual information between interactions. This limitation introduces significant challenges for reliable LLM-based multi-step GUI testing, where sustained contextual coherence across iterative interactions is essential. Without persistent context, the testing process risks producing inconsistent or redundant behaviors. To overcome this issue, \toolname introduces a structured \cm inspired by human memory theory \cite{baddeley1992working}. The \cm acts as a central repository for storing and managing diverse types of information generated during testing, thereby supporting reasoning, decision-making, and feedback across agents. Specifically, it is divided into three components: long-term memory, which stores static information such as device configuration, app metadata, and testing scenarios; working memory, which maintains dynamic information including current testing goals, executed actions, and agent interactions; and short-term memory, which records transient GUI states and recognition results. This hierarchical organization enables seamless information flow across iterations, ensuring that testing decisions remain contextually grounded and progressively adaptive.

\textbf{Long-term Memory.} 
This memory stores static and global information that remains consistent throughout the entire testing campaign. It includes details about the testing environment (\eg device information, operating system version), app metadata (\eg app name, package identifier, version), and descriptions of the testing scenarios. These data are organized in a structured format and provide stable background knowledge for the LLM-based agents. The long-term memory ensures that agents can interpret GUI semantics and testing scenario goals in the correct app context, eliminating the need for redundant reinitialization across iterations or app sessions.

\textbf{Working Memory.} 
The working memory retains dynamic but persistent information relevant to the current testing session. It contains the target testing scenario, the current app under test, the history of executed operations, intermediate outcomes, and the recent interaction history with the LLMs. Unlike long-term memory, working memory evolves continuously as the test progresses, reflecting changes in GUI states and user-like actions. Each agent accesses this memory to maintain consistency and traceability. For example, the \de consults it to decide the next operation based on past actions, while the \su refers to it when verifying whether a sequence of actions aligns with the intended testing scenario. The structured storage enables the system to efficiently retrieve relevant records for reasoning and prompt generation.

\textbf{Short-term Memory.} 
The short-term memory captures transient and immediate data associated with the most recent interactions. It stores the current GUI state, the preceding GUI state, and the results of widget recognition performed by the \ob. Such information is frequently updated as the app interface evolves during testing. The app states are represented primarily as GUI screenshots, augmented with detected widget attributes and layout structures. The short-term memory enables quick contextual reasoning, for example, allowing the \su to compare the current GUI against the previous one to determine whether the intended operation succeeded or whether a new state transition occurred.

To improve efficiency and prevent redundant exploration, \toolname employs a GUI state comparison mechanism within the short-term memory. Two GUI states are considered identical if their visual and structural representations differ only in trivial details (\eg minor animations, timestamps, or non-functional layout shifts). This comparison leverages perceptual hashing and structural similarity metrics to identify equivalent states. When an identical state is detected, the system skips re-exploration and updates only the relevant context, significantly reducing unnecessary operations and improving testing efficiency.

Overall, the \cm forms the cognitive foundation of \toolname iterative process. It is dynamically updated during widget recognition, action execution, and test recording, while decision-making and validation stages rely on its stored content to formulate prompts, reason about app behavior, and ensure consistency across testing iterations. This design enables the agents in \toolname to emulate human-like continuity and memory-driven reasoning during automated scenario-guided GUI testing.

\subsection{Observer}

\begin{figure}[!htbp]
\centering
\includegraphics[width=0.7\linewidth]{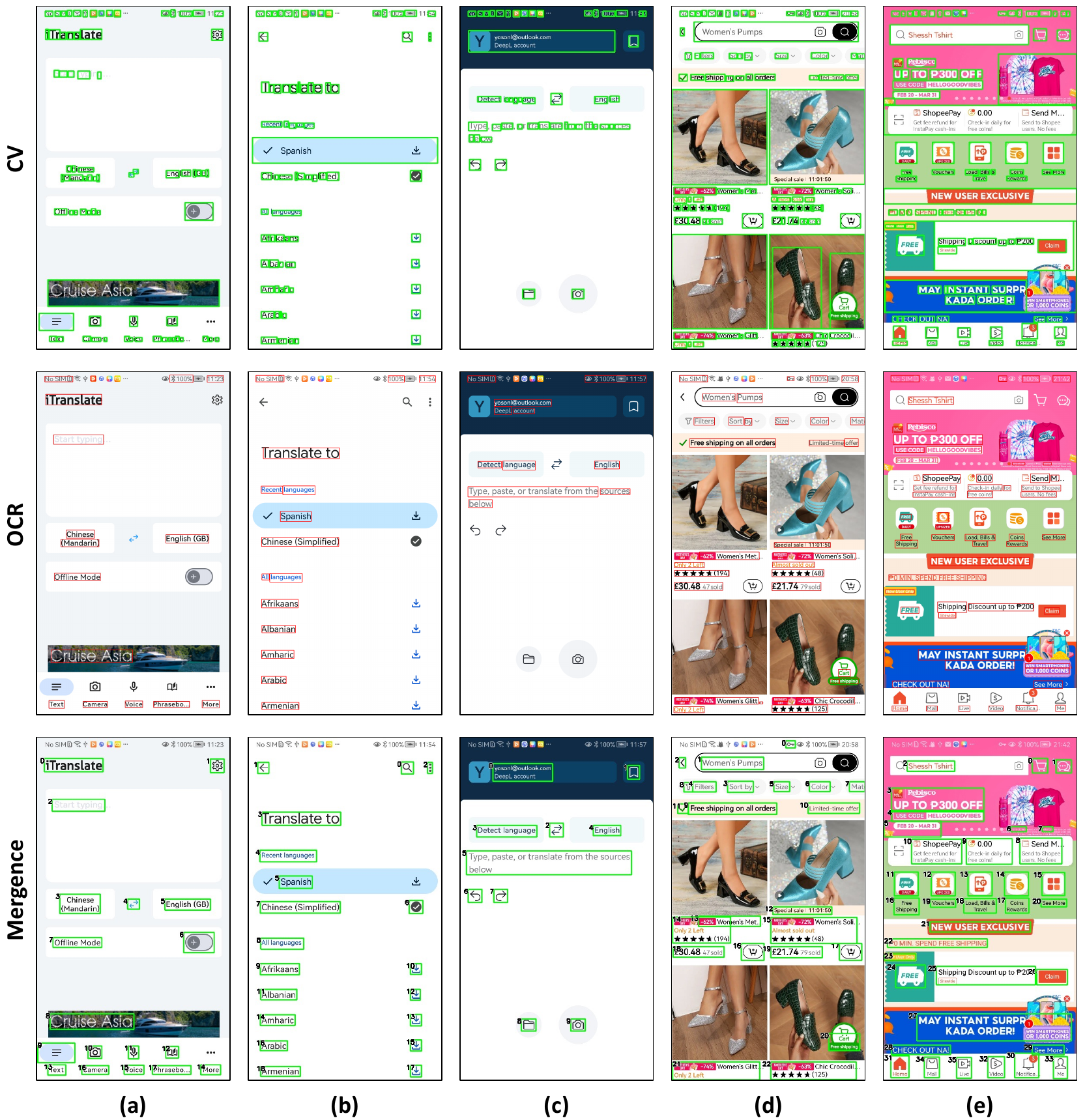}
\caption{Result Mergence Example}
\label{fig:merge}
\end{figure}

The \ob is responsible for perceiving the app's GUI and constructing a structured representation of its current state to support subsequent decision-making and testing operations. However, accurately understanding GUI states presents a dual challenge: traditional computer vision (CV) techniques offer precise pixel-level detection but lack semantic comprehension, while large language models (LLMs) possess strong reasoning abilities yet struggle with fine-grained localization. To overcome this limitation, \toolname adopts a hybrid perception design that combines traditional CV algorithms with multi-modal LLM reasoning. The CV component extracts and localizes widgets through visual analysis, while the LLM interprets their semantic roles and contextual relationships within the interface. This integration enables the \ob to achieve both spatial precision and semantic depth, producing a comprehensive, human-like understanding of the GUI state essential for accurate, scenario-driven test generation.

The widget recognition process primarily targets GUI screenshots and proceeds in three main stages. First, CV algorithms are employed to identify potential GUI widgets within the GUI image. Specifically, we adopt a UIED-style \cite{xie2020uied} pipeline combining edge detection, contour analysis, and component grouping to identify potential GUI widgets. The rationale for using such traditional CV techniques is their determinism and pixel-level precision. These methods are effective in extracting structural layout information without requiring access to source code or layout files, which ensures platform independence and applicability to hybrid or dynamically rendered interfaces. Second, optical character recognition (OCR) algorithms are applied to extract textual elements from the same screenshot, capturing words, phrases, and textual labels. These two stages provide the foundational visual and textual components for the subsequent fusion stage. Finally, the recognition results are integrated and refined to generate the final widget set, ensuring that both visual completeness and semantic coherence are preserved.

During this fusion and filtering stage (examples shown in \figref{fig:merge}), several refinements are applied. We merge adjacent text fragments based on spatial proximity to form coherent textual widgets and exclude unreasonable GUI detections, such as duplicated text regions, status-bar icons (\eg battery indicators or signal symbols), or objects that are excessively large (encompassing multiple widgets) or too small (noise artifacts). Additionally, related graphical and textual widgets are merged into unified widget entities when they jointly represent the same user-interaction target (\eg an icon with a label underneath). The merging is guided by a predefined distance threshold, following prior empirical settings \cite{yu2024effective, yu2024practical}. Concerning threshold selection and its impact on generalizability, thresholds primarily influence spatial proximity decisions during widget mergence. If the threshold is too large, independent interactive elements may be combined, reducing selection precision. If too small, logically unified widgets may be fragmented, increasing decision ambiguity. However, mobile UI design conventions typically follow standardized layout spacing and alignment patterns. Empirical observation across diverse applications suggests that a fixed, moderately calibrated threshold performs robustly for mainstream UI designs. Therefore, while threshold values affect fine-grained detection behavior, they do not fundamentally limit the applicability of the framework to different apps. The approach remains general because it operates on visual layout structures rather than app-specific hard-coded rules.

It is important to note that \toolname adopts a vision-based recognition strategy rather than relying solely on extracting widget information from GUI layout files. This decision is motivated by practical limitations: widgets embedded in \texttt{Canvas} objects or hybrid H5 pages may be inaccessible through layout parsing \cite{yu2025vision, yu2024practical, yu2024effective}. Directly identifying widgets from rendered screenshots ensures a ``what-you-see-is-what-you-get'' representation of the GUI, aligning the perceived interface with the user's actual view and preserving fidelity in testing.

After obtaining the final widget recognition results, the \ob invokes a multi-modal LLM instance to perform semantic interpretation of the GUI state. The LLM analyzes visual layouts and textual cues to infer the roles and relationships of widgets (\eg recognizing navigation buttons, input fields, or functional icons). This semantic understanding enriches the purely structural detection results from the CV stage, enabling subsequent agents (\de, \su) to reason about GUI intent and business logic during scenario-guided testing.

Finally, all recognized widgets are visualized by overlaying bounding boxes and IDs on the original screenshot. This annotated image, along with the structured widget list and the raw GUI image, is stored in the short-term memory of \toolname. This visual-structural representation not only facilitates cross-agent communication but also serves as a crucial input for LLM-based reasoning and action planning in later testing stages. The motivation of using annotated images is grounded in the limitations of multi-modal LLMs in pixel-level localization \cite{yang2023set}. While LLMs exhibit strong semantic reasoning ability, they are less reliable in predicting exact coordinate-level actions directly from raw screenshots. Annotated images serve as a visual grounding mechanism that externalizes candidate widgets with bounding boxes and identifiers. This reduces the reasoning burden from coordinate inference to semantic selection. The LLM selects a widget at the conceptual level, and the system maps that selection to precise coordinates obtained from CV detection. This separation improves reliability and execution accuracy while preserving semantic reasoning capability.

\subsection{Decider}

The \de is responsible for determining the next operation during scenario-guided GUI test generation, guiding the testing process toward fulfilling the target testing scenario. In \toolname, test generation is modeled as an iterative decision-making process that begins from the app's initial state and advances through a sequence of context-aware actions until the testing goal is achieved. However, this process presents two key challenges: first, accurately reasoning about the next action requires understanding both the current GUI state and the overall business logic of the target testing scenario; second, LLMs, while capable of semantic reasoning, may struggle to align their decisions with precise, executable GUI operations. To address these issues, \toolname employs an LLM-based decision mechanism that transforms GUI understanding and testing scenario semantics into concrete, stepwise operations. Each iteration evaluates the current app state, the history of executed actions, and the predefined testing scenario context to infer the most appropriate next operation through LLM reasoning. This design enables the \de to bridge abstract testing scenario understanding and concrete GUI interactions, ensuring that every decision aligns logically with the testing scenario progression and contributes effectively to testing scenario completion. Every action decision is derived as an inference mapping generated by an LLM, formulated as follows:

\begin{equation}
(T, S, L_{op}, P_{curr}, W_{curr}) \stackrel{LLM} \longrightarrow op
\end{equation}

where $T$ denotes the action type set, $S$ denotes the target testing scenario, $L_{op}$ denotes the list of previously executed actions, $P_{curr}$ denotes the current GUI, and $W_{curr}$ denotes the widget set of the current GUI.

\textbf{Structure of Target Testing Scenarios.}
A target testing scenario $S$ represents a functional goal expressed through natural language, describing the intended user operation sequence or outcome. Each testing scenario is composed of (1) a concise textual description (\eg ``log into the app using valid credentials'') and (2) optional contextual information such as the expected starting page or relevant GUI screenshots. This natural language definition allows the LLM-based \de to interpret the high-level intent and decompose it into a sequence of coherent GUI actions. The flexibility of testing scenario descriptions enables the framework to generalize across diverse apps while still grounding reasoning in the specific GUI context provided by the \ob.

Compared to traditional LLMs, multi-modal LLMs can directly process GUI screenshots, enabling visual-semantic reasoning and action planning. Thus, \toolname employs raw GUI images and utilizes multi-modal LLMs to make visual decisions directly. However, this design introduces a cognition–execution gap: while LLMs excel at reasoning about what should be done, they lack the ability to locate precise pixel-level widget positions required for action execution \cite{liu2024testing}. To address this, \toolname integrates multi-modal reasoning with traditional CV-based widget localization. The decision process is thus divided into two stages, logical decision-making and widget localization, bridging high-level reasoning and low-level action feasibility.

\begin{figure}[!htbp]
\centering
\includegraphics[width=\linewidth]{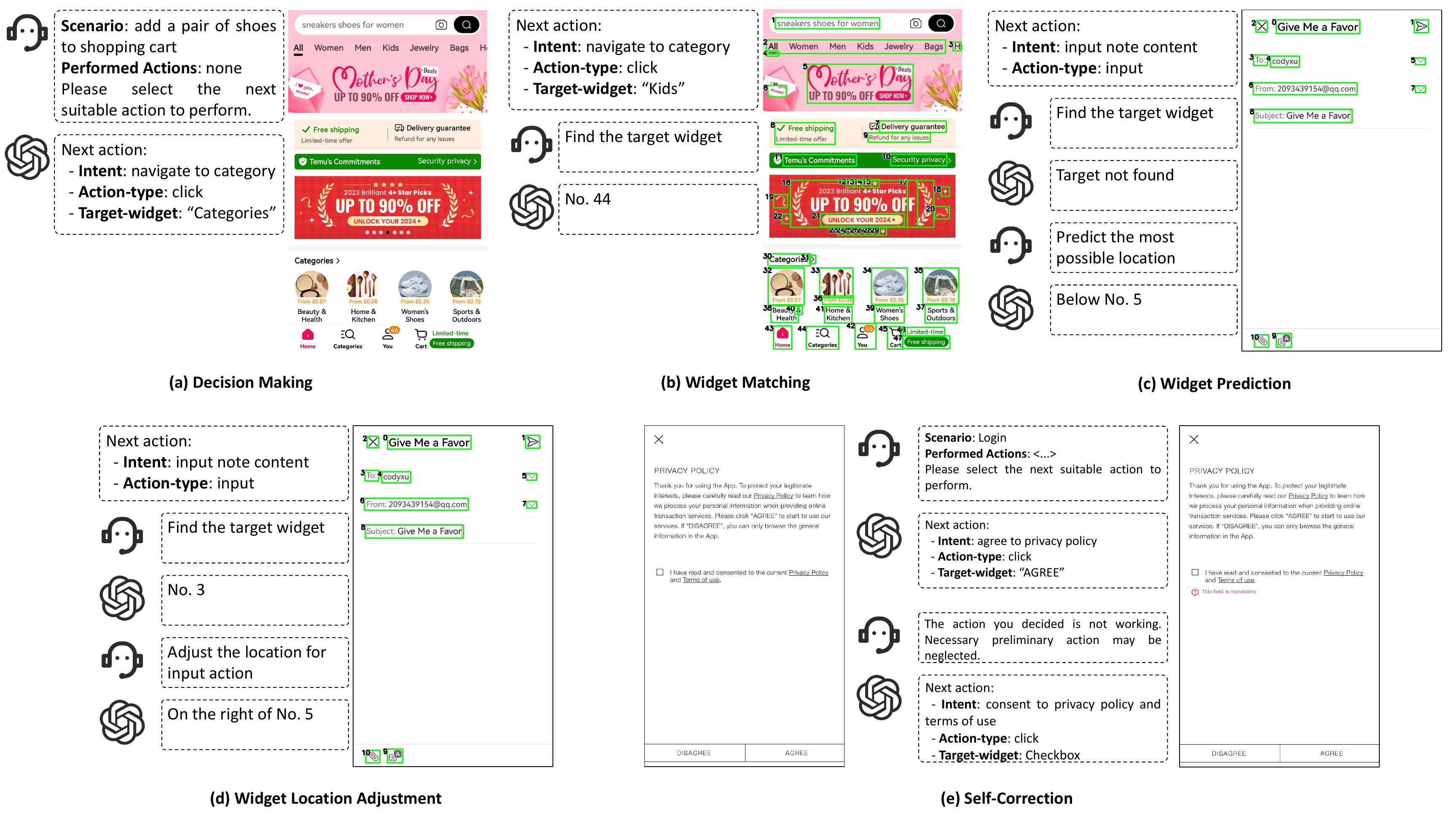}
\caption{LLM Interaction of \de}
\label{fig:decider}
\end{figure}

\subsubsection{Logical Decision-making}

The logical decision-making stage leverages the reasoning capabilities of multi-modal LLMs to analyze GUI images, interpret the current state, and decide the next logical operation. By combining the action type set $T$, the target testing scenario $S$, and the list of previously executed actions $L_{op}$, the LLM determines the next appropriate action $op'$:

\begin{equation}
(T, S, L_{op}, P_{curr}) \stackrel{LLM} \longrightarrow op'
\end{equation}

The decision output $op'$ contains: (1) the type of action (\eg click, input, scroll), (2) a natural language description of the action intent, (3) a description of the target widget, (4) the input text (for input actions), and (5) the scroll direction (for scroll actions). For efficiency, \toolname merges the implicit ``focus click'' operation required for input actions into a single composite action, avoiding redundant planning steps. \figref{fig:decider}(a) illustrates this process.

\textbf{Handling Complex Actions.}
Some testing scenarios require executing compound actions, \ie coordinated sequences of interactions across multiple widgets (\eg entering a username and password, then clicking a login button). The \de manages these cases through dynamic action composition, in which the LLM decomposes the high-level intent into smaller, coherent sub-operations that maintain contextual continuity. Each sub-action is generated sequentially, guided by the testing scenario context and previous decisions stored in the \cm. This design enables flexible reasoning over multi-step workflows without requiring predefined action templates.

\textbf{Text Value Generation.}
For input actions, the text content is dynamically produced according to the testing scenario context. The \de first interprets semantic cues from the testing scenario description (\eg ``enter username'', select a realistic username). If the testing scenario provides explicit textual values, these are directly used; otherwise, the LLM generates contextually plausible text based on general knowledge or app-specific hints extracted from the GUI (\eg placeholders, labels). This approach ensures that input data remain both semantically valid and consistent with user-intent reasoning, enabling realistic test execution.

\subsubsection{Target Widget Location}

After logical decision-making, the action intent and the target description are available, but precise pixel-level widget coordinates are still unknown. To enable actionable execution, the \de performs widget localization, refining the decision into an executable operation:

\begin{equation}
(op', W_{curr}) \stackrel{LLM} \longrightarrow op
\end{equation}

\textbf{Widget Matching.}
Widget matching aligns the abstract target description from the decision stage with concrete GUI elements identified by the \ob. To ensure contextual coherence, \toolname provides the LLM with visualized widget recognition results (annotated bounding boxes and IDs) as prompts, enabling it to correlate semantic intent with the correct widget on the interface. This direct comparison between annotated and raw screenshots ensures accurate widget identification, as shown in \figref{fig:decider}(b).

\textbf{Widget Prediction.}
In cases where certain widgets are visually indistinct or missing from recognition outputs (\eg due to blurred boundaries or dynamic layouts), \toolname employs a widget prediction mechanism. Using the relative positions and spatial relationships of nearby recognized widgets, the framework estimates the most probable location of the missing target, generating a ``virtual widget'' to maintain logical flow. This ensures robustness against incomplete GUI detection, as illustrated in \figref{fig:decider}(c).

\textbf{Widget Location Adjustment.}
Before executing input actions, \toolname adjusts click positions to match functional regions. For example, if the target ``To:'' label is identified during email composition, the system shifts the click focus to the adjacent text box, which is the actual interactive element. This adjustment guarantees that interactions correspond to functional widgets rather than purely decorative or descriptive elements \cite{yu2024practical}. \figref{fig:decider}(d) illustrates such adjustment.

\subsubsection{Self-Correction}

When a recent decision fails state transition verification, indicating that the intended effect was not achieved, the self-correction mechanism is triggered. This mechanism includes two steps: cause analysis and decision adjustment. The cause analysis identifies one of three common error types: (1) logical decisions were correct, but widget localization failed; (2) logical decisions were reasonable, but required preconditions (\eg navigation steps) were missed; or (3) the logical decision itself deviated from testing scenario intent.

Based on the identified cause, the correction is applied in the corresponding dialogue context. For widget localization errors, the context of the localization prompt is reused to request revised reasoning. For logical or contextual errors, the dialogue from the decision-making phase is revisited, and the LLM is prompted to produce an adjusted plan. Through this iterative feedback loop, \toolname ensures that each decision advances testing progress coherently and aligns with the target testing scenario. An example of this self-correction process is shown in \figref{fig:decider}(e).

\subsection{Executor \& Recorder}

The \ex and \re in \toolname are designed to ensure the stability, traceability, and reproducibility of test execution. These modules are responsible for translating the high-level action plans generated by preceding agents into concrete GUI interactions and for recording the corresponding results and runtime information. This separation between reasoning (handled by LLM-based agents) and execution (handled by rule-based modules) prevents error propagation from uncertain LLM outputs and guarantees consistent interaction with the app under test.

\toolname supports a wide range of common GUI interaction operations, including click, input, scroll, back, and others, which together cover the majority of user actions encountered during mobile app testing. The execution process begins with the \ex parsing the structured action decision produced by the \de. It then invokes system-level interfaces (\eg Android ADB commands) to perform the specified operation on the device. After execution, the \re logs the outcomes, including executed actions, GUI states, and system responses, into the working memory of the \cm, ensuring that subsequent iterations have full context for reasoning and verification. Additionally, the \re monitors runtime logs to detect anomalies such as app crashes or unhandled exceptions, which are recorded as potential defects during exploration.

Each operation type may require different parameters during the parsing phase. For click actions, the target widget must be specified. Input operations require both the target widget and the text content, as well as information about the input's positional relationship to the widget. Scroll operations demand a defined scroll direction and, optionally, a target region (defaulting to the entire screen if unspecified). Notably, input operations are executed as a two-step process: first, performing a click to focus on the input field, followed by the text entry. The back operation, by contrast, requires no additional context parameters.

The script construction process begins after each action has been executed and verified. During execution, the \ex performs concrete GUI operations via system-level interfaces such as ADB, while the \re captures all relevant contextual information associated with that operation. This includes the high-level operation intent generated by the \de, the concrete action type, the resolved target widget attributes, the exact execution parameters, and the resulting GUI state. Rather than storing only raw commands, the framework records semantically enriched action entries to preserve traceability between testing scenario intent and low-level execution.

Each recorded action is then transformed into a structured representation. Specifically, the script entry includes the operation type such as click, input, scroll, or back; the textual description of the operation intent; the target widget's semantic attributes including widget text, bounding box coordinates, and widget description; the file path of the associated pre- and post-action screenshots; and the exact ADB command that was executed. For input actions, the script explicitly stores the input text and the focus activation step to ensure correct replay semantics. This structure ensures that the action is both human-readable and machine-executable.

At the completion of a testing scenario, the accumulated action entries are serialized into a JSON-formatted script. Device-level metadata such as screen resolution is also included to guarantee correct coordinate interpretation during replay. The resulting JSON file therefore serves as a deterministic replay script that can be executed independently of the original LLM reasoning process. This design intentionally decouples exploratory decision-making from final artifact generation. The LLM agents are responsible for intelligent exploration and reasoning, whereas the constructed script represents a stable, reproducible outcome suitable for regression testing, debugging, and empirical evaluation.

The intuition behind this construction strategy is to ensure that LLM-based reasoning does not remain ephemeral. Instead of relying on runtime LLM prompts for future reuse, the framework externalizes the generated behavior into an explicit and inspectable test artifact. This bridges agentic AI exploration with conventional software testing workflows, allowing generated tests to be replayed, shared, and validated without re-invoking the LLM.

By implementing execution and recording as deterministic processes, \toolname ensures that every GUI interaction is repeatable and verifiable, forming a reliable foundation for multi-agent collaboration. This design not only strengthens traceability across test iterations but also facilitates accurate post-analysis of failures and system behaviors, supporting robust and transparent scenario-guided GUI testing.

\subsection{Supervisor}

The \su is responsible for verifying the app state after each executed action, ensuring that the operation outcomes align with expectations before the next decision is made. Accurate verification is essential to maintain consistency between testing scenario intent and actual app behavior during iterative test generation. However, GUI state validation in mobile apps is inherently challenging due to dynamic page transitions, asynchronous loading behaviors, and diverse feedback mechanisms. Without robust verification, minor visual delays or hidden transitions may lead to misjudged states, erroneous decisions, or premature testing scenario termination. To address these challenges, the \su employs a three-stage verification mechanism consisting of real-time loading verification, state transition verification, and test completion verification. This design allows the system to first ensure the app has reached a stable state, then confirm that state changes correspond to the expected outcomes, and finally determine whether the target testing scenario has been completed. By continuously validating and correcting deviations, the \su provides a feedback loop that enhances system reliability, ensuring accurate and traceable progress throughout scenario-guided test execution.

\begin{figure}[!htbp]
\centering
\includegraphics[width=\linewidth]{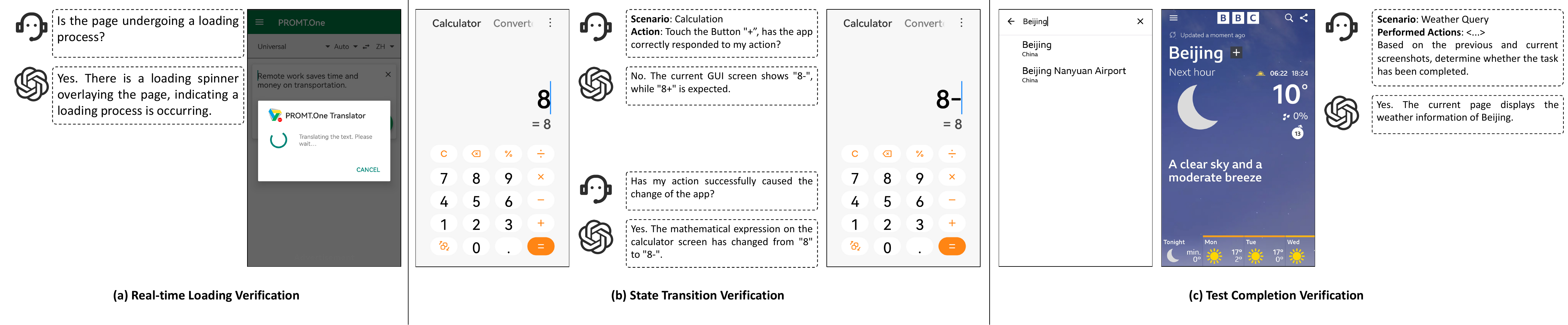}
\caption{LLM Interaction of \su}
\label{fig:supervisor}
\end{figure}

\textbf{Real-time Loading Verification.} 
The execution of a test operation may temporarily place the app into an unstable state, such as a loading or transition phase triggered by a network request. In such cases, premature decision-making may lead to misinterpretation of the app state or the early triggering of self-correction. While a fixed waiting time could mitigate this issue, it is difficult to generalize and often harms testing efficiency. Instead, \toolname leverages the visual understanding capability of a multi-modal LLM to dynamically determine whether the app has completed loading. By analyzing the current GUI image, the \su detects loading indicators or transitional states and adjusts the waiting duration adaptively. This adaptive waiting strategy improves both the reliability and efficiency of the testing process. An example of this is shown in \figref{fig:supervisor}(a).

\textbf{State Transition Verification.} 
After ensuring that the app is stable, the next step is to verify whether the most recent operation has produced the intended state transition. The \su leverages a multi-modal LLM to analyze and compare the current GUI state $P_{curr}$ with the previous GUI state $P_{prev}$, conditioned on the most recent operation $op_{latest}$ and the target testing scenario $S$. Specifically, both screenshots, together with their annotated widget structures, are provided as joint visual inputs to the LLM, which performs semantic-level comparison rather than raw pixel differencing. The model reasons about changes in visible widgets, textual content, layout structure, and interaction-relevant elements to determine whether a meaningful state transition consistent with $op_{latest}$ has occurred under scenario $S$. The process is expressed by:

\begin{equation}
(S, op_{latest}, P_{curr}, P_{prev}) \stackrel{LLM} \longrightarrow result
\end{equation}

If the verification result indicates failure, an additional comparison between $P_{curr}$ and $P_{prev}$ is conducted to determine whether any actual change has occurred:

\begin{equation}
(P_{curr}, P_{prev}) \stackrel{LLM} \longrightarrow result
\end{equation}

This step distinguishes between legitimate state transitions and irrelevant or automatic changes, such as clock updates, animated banners, or transient notifications. When no observable or meaningful effect is detected, \toolname categorizes the operation as a non-effective action. In such cases, the system initiates a self-correction mechanism that re-evaluates the previous decision and generates an alternative operation targeting the same functional goal.

In situations where the executed operation produces unintended side effects, for example, navigating to an incorrect page or triggering an unwanted dialog, the \su identifies the deviation through semantic comparison of the GUI state. The framework then performs a contextual rollback: it executes a predefined reverse action (\eg a ``back'' operation) or reverts to the last valid state stored in the \cm. \figref{fig:supervisor}(b) provides an example of this verification process. This design ensures that the exploration process remains consistent with the target testing scenario and prevents error accumulation in subsequent iterations. When the GUI effect is brief or difficult to capture (\eg a short-lived popup), the \su records a sequence of screenshots during the verification window and uses the LLM to detect temporal changes, ensuring that short-duration effects are properly recognized. This process is automatically triggered by the \su when verification uncertainty is detected. After an action execution, if the immediate post-action screenshot does not provide sufficient evidence of state transition, or if the executed action is likely to produce transient visual feedback such as popups or animations, the system programmatically enters a short observation window. During this window, multiple screenshots are captured at predefined intervals. This strategy allows the framework to detect ephemeral state changes that may not persist long enough for a single-frame capture. The trigger condition is based on execution context and verification confidence rather than manual specification, ensuring fully automated behavior.

\textbf{Test Completion Verification.}
The final stage determines whether the testing process has reached the expected endpoint for the current testing scenario. Because the endpoint cannot be predetermined, the \su performs real-time evaluation during exploration to detect completion conditions. Leveraging a multi-modal LLM, it integrates information from the target testing scenario $S$, the list of executed operations $L_{op}$, and both the current and previous GUI screenshots ($P_{curr}$ and $P_{prev}$) to assess progress:

\begin{equation}
(S, L_{op}, P_{curr}, P_{prev}) \stackrel{LLM} \longrightarrow result
\end{equation}

If the LLM determines that the functional goal of $S$ has been achieved or the testing path sufficiently covers the intended testing scenario, the testing task is automatically terminated. Conversely, if an incomplete state is detected, the system continues the exploration process, guided by the \de. This ensures that \toolname terminates testing precisely when the target functionality has been comprehensively validated. An illustrative example is shown in \figref{fig:supervisor}(c).

Through this multi-level verification and correction mechanism, the \su guarantees that the app state transitions are accurately recognized, transient or spurious effects are properly handled, and unintended side effects are mitigated. This design ensures robust, context-aware supervision throughout the entire scenario-guided testing process.

\toolname does not generate traditional assertion-based test oracles. The framework is designed for scenario-driven test generation, where the primary objective is to guide the system toward completing predefined high-level testing scenarios rather than producing explicit assertions such as equality checks or expected-value comparisons. Instead of static oracles, \toolname adopts a dynamic verification strategy implemented by the Supervisor.

The \su ensures that the generated test remains aligned with the intended testing scenario throughout the iterative process. After each executed action, it evaluates the resulting GUI state and determines whether the system has progressed consistently toward the testing scenario goal. When deviations occur due to incorrect widget selection, misunderstanding of the application state, or errors in action logic, the \su detects the inconsistency, triggers backtracking when necessary, and supports re-decision by the Decider to redirect the testing process toward the expected testing scenario path. In this sense, the \su functions as a scenario-level consistency validator rather than a traditional oracle generator.

\section{Empirical Evaluation}

\subsection{Experiment Settings}

To comprehensively evaluate the performance and practicality of \toolname, we design five research questions (RQs) that assess its effectiveness, efficiency, and real-world applicability:

\begin{itemize}

\item \textbf{RQ1}: How effective is the scenario-guided test generation of \toolname?
\item \textbf{RQ2}: How effective is the bug detection capability of \toolname?
\item \textbf{RQ3}: How effective is the widget localization of \toolname?
\item \textbf{RQ4}: How effective is the logical decision-making of \toolname?
\item \textbf{RQ5}: How efficient is the scenario-guided test generation of \toolname?
\end{itemize}

These research questions jointly provide a comprehensive assessment of \toolname from multiple perspectives. Specifically, RQ1 investigates the overall quality of the generated tests, verifying the ability of \toolname to produce meaningful and scenario-aligned testing sequences. RQ2 assesses the framework's effectiveness in identifying real-world software bugs, demonstrating its practical utility. RQ3 measures the precision of widget localization conducted by the \ob, focusing on accurate visual grounding within GUI states. RQ4 evaluates the accuracy of logical decision-making by the \de, assessing the LLM's ability to produce semantically valid and contextually relevant actions. Finally, RQ5 examines the computational and temporal efficiency of test generation. Together, these RQs provide a multi-dimensional understanding of the capability and robustness of \toolname.

\textbf{LLM Configuration.}
We employ the GPT-4V model \cite{openai2024gpt4} as the LLM instance in \toolname, leveraging its advanced visual understanding capabilities for reasoning about GUI layouts and testing scenario semantics. The model is accessed through the \texttt{gpt-4-vision-preview} API, which enables direct visual processing of app screenshots. GPT-4V is chosen for its strong performance in multi-modal reasoning tasks, its ability to interpret complex visual contexts in combination with textual prompts, and its demonstrated robustness in structured decision-making. These characteristics make it particularly suitable for GUI testing, where accurate comprehension of widget semantics, layout hierarchy, and visual cues is essential for generating meaningful scenario-driven actions. The parameter \texttt{temperature} is fixed at 0 to promote determinism and minimize stochastic response variation across runs, ensuring consistency in logical reasoning and decision outcomes while maintaining fairness and reproducibility in testing results.

\begin{table}[!htbp]
\centering
\caption{Experiment Subject: Scenarios and Apps}
\scalebox{1}{
\begin{tabular}{llc}
\toprule
Scenario    & Description & \# Task \\ \midrule
Email       & Create, edit and send an email                   & 6  \\
Music       & Search, choose and play the music                & 9  \\
Note        & Create, edit and save the note                   & 16 \\
Photo       & Take a photo                                     & 9  \\
Translation & Choose a language, edit the text, and translate  & 13 \\
Shopping    & Search for an item and save to the shopping cart & 6  \\
Login       & input the required token to login                & 7  \\
Calculation & Math calculation with calculator                 & 13 \\
Weather     & Query the weather of specific areas              & 10 \\
Alarm-clock & Create and configure the alarm-clock             & 10 \\ \bottomrule
\end{tabular}}
\label{tab:scenario}
\end{table}

\textbf{Testing Scenario and App Selection.}
To establish realistic and representative testing targets, we conduct an investigation of the Google Play Store to identify the most prevalent app categories based on download volume and the number of available apps. From these categories, we extract the most frequent and essential business functionalities as candidate testing scenarios. The testing scenarios are created by three experienced software testing researchers, each with over five years of experience in mobile app testing and GUI automation. To reduce subjectivity, the researchers follow a structured testing scenario definition protocol: (1) each testing scenario is first described in natural language to represent a specific end-user goal; (2) all draft testing scenarios are reviewed by the research team to ensure coverage across different functional domains; and (3) overlapping or ambiguous testing scenarios are merged or refined through discussion. This collaborative design process ensures that the testing scenarios are representative of realistic user interactions while minimizing developer bias. The scenarios are not artificially engineered test scripts. They originate from common user goals that naturally exist across mobile applications, such as sending an email, completing a purchase, or logging into an account. Our contribution lies in abstracting these user goals at a semantic level and formalizing them as structured inputs to the framework. We do not manually encode detailed procedural steps; instead, the agentic system autonomously derives operational sequences. Therefore, the manual effort is limited to defining high-level scenario descriptions, which typically requires only a short textual specification.

As summarized in \tabref{tab:scenario}, ten representative testing scenarios are finalized, covering a range of functional complexities. The first column lists the testing scenario names, the second provides textual descriptions, and the third shows the number of apps associated with each. Based on these testing scenarios, we select a total of 92 apps (with 7 used in two testing scenarios) and construct 99 tasks (\ie one specific testing scenario on one specific app). The full list of apps is available on our project website. To enhance coverage and realism, we include both open-source and commercial apps collected from multiple major platforms, including the Google Play Store and the Huawei AppGallery. Since our LLM performs best on English-language content, only apps with English interfaces are selected to eliminate confounding effects due to multilingual GUIs. The chosen apps span a wide range of functionalities and interaction complexities, ensuring the diversity and validity of our evaluation.

\textbf{Baselines.}
We employ several baselines to facilitate fair and meaningful comparison across research questions. For RQ1 and RQ5, we compare against three state-of-the-art scenario-guided or context-aware testing frameworks: GPTDroid \cite{liu2024make}, ScenTest \cite{yu2024practical}, and ASOT\footnote{Not an official name from the authors.} \cite{su2025automated}. These baselines represent recent advances in scenario-guided test generation. For RQ2, which examines bug detection effectiveness, we benchmark against four representative automated GUI testing tools: Monkey \cite{monkey} (random-based), Stoat \cite{su2017guided} (model-based), PIRLTest \cite{yu2024effective} (learning-based), and ScenTest \cite{yu2024practical} (scenario-based), covering a spectrum of exploration strategies. This selection ensures that performance is compared across both traditional and modern paradigms in automated GUI testing. For RQ3 and RQ4, which focus on internal mechanisms of \toolname, we use quantitative evaluations without external baselines. 

\subsection{RQ1: Scenario-Guided Test Generation Capability}

In RQ1, we evaluate the overall capability of \toolname to automatically generate GUI tests that accurately and reliably complete predefined testing scenarios. This assessment aims to verify whether \toolname can effectively understand, plan, and execute multi-step scenario-specific testing processes, thereby demonstrating its practical usability and reliability.

To characterize the outcome of each test generation task, we classify results into four cases:

\begin{itemize}
\item Case 1 ($c_1$): The test generation completes successfully and fully covers the target testing scenario.
\item Case 2 ($c_2$): The test generation terminates abnormally but still fully covers the target testing scenario.
\item Case 3 ($c_3$): The test generation completes successfully but fails to fully cover the target testing scenario.
\item Case 4 ($c_4$): The test generation terminates abnormally and fails to fully cover the target testing scenario.
\end{itemize}

Based on these classifications, we define two quantitative evaluation metrics: \textbf{Coverage Rate} ($C$) and \textbf{Success Rate} ($S$). $C$ measures the proportion of test generations that achieve complete testing scenario coverage, regardless of termination status, while $S$ evaluates the proportion of tests that both complete successfully and achieve full coverage. These metrics are calculated as follows:

\begin{align}
C=\frac{\vert c_1 \vert+\vert c_2 \vert}{\vert c_1 \vert + \vert c_2 \vert + \vert c_3 \vert + \vert c_4 \vert} \times 100\%, \quad
S=\frac{\vert c_1 \vert}{\vert c_1 \vert + \vert c_2 \vert + \vert c_3 \vert + \vert c_4 \vert}\times 100\%
\end{align}

where $\lvert c_i\rvert$ denotes the number of instances in case $c_i$.

\begin{table}[!htbp]
\centering
\caption{Result of RQ1: Scenario-based Test Generation Capability (in total of 99 tasks)}
\begin{tabular}{c|ccc|ccc|c}
\toprule
 & GPTDroid & ScenTest & ASOT & Mobile-Agent-v3 & AppAgent & AutoDroid & ScenGen \\
\midrule
$\vert c_1\vert$ & 68 & 69 & 73 & 75 & 78 & 80 & 84 \\
$\vert c_2\vert$ & 6  & 8  & 11 & 8  & 5  & 4  & 2  \\
$\vert c_3\vert$ & 9  & 11 & 8  & 9  & 7  & 8  & 7  \\
$\vert c_4\vert$ & 16 & 11 & 7  & 7  & 9  & 7  & 6  \\
\midrule
C & 74.75\% & 77.78\% & 84.85\% & 83.84\% & 83.84\% & 84.85\% & 86.87\% \\
S & 68.69\% & 69.70\% & 73.74\% & 75.76\% & 78.79\% & 80.81\% & 84.85\% \\
\bottomrule
\end{tabular}
\label{tab:rq1new}
\end{table}

The experimental results are summarized in \tabref{tab:rq1new}. Across 99 total test generation tasks, \toolname achieves the highest Coverage Rate (86.87\%) and Success Rate (84.85\%) among all compared methods. GPTDroid, ScenTest, and ASOT obtain Coverage Rates of 74.75\%, 77.78\%, and 84.85\%, respectively, and Success Rates of 68.69\%, 69.70\%, and 73.74\%. This comparison demonstrates that \toolname surpasses all baselines in both comprehensive coverage and stable completion across diverse testing scenarios. Specifically, GPTDroid and ScenTest, though both scenario-oriented, rely primarily on text-based reasoning or pre-defined rules for test planning, leading to higher rates of premature or incomplete executions. ASOT achieves relatively strong results due to its adaptive testing scenario graph modeling, yet it lacks the semantic flexibility to handle unexpected GUI transitions or dynamically generated elements. In contrast, \toolname benefits from its LLM-based multi-agent design and context memory integration, allowing it to interpret GUI semantics and dynamically adapt decision-making to interface feedback.

Moreover, \toolname exhibits notably fewer abnormal terminations (\textit{Case 2} and \textit{Case 4}), indicating that the combination of multi-modal LLM reasoning and the feedback mechanism of \su enables more stable iterative exploration. The smaller proportion of incomplete coverage cases (\textit{Case 3}) also demonstrates that the testing scenario understanding of \toolname allows it to maintain logical coherence throughout test generation, avoiding early exits that commonly affect baseline tools. From the 99 test generation tasks, \toolname records 84, 2, 7, and 6 instances of $c_1$, $c_2$, $c_3$, and $c_4$, respectively. These results highlight that even in complex testing environments involving different app architectures and UI layouts, \toolname can consistently generate tests that fully achieve the testing scenario objectives in most cases. The high coverage and success rates reflect both the reliability and generalizability of the approach across different domains.

Despite its strong overall performance, we conduct a detailed analysis of the failure cases (from $c_2$, $c_3$, and $c_4$) to identify improvement opportunities. The main causes of failure include premature termination, inaccurate completion detection, adaptation issues to app-specific designs, imprecise widget localization, and insufficient flexibility in handling unconventional user flows. Among them, Premature Termination accounts for 6 cases, Inaccurate Completion Detection accounts for 3 cases, Widget Localization Errors account for 3 cases, App-specific Design Variations account for 2 cases, and Handling Unconventional Processes accounts for 1 case. Premature Termination is therefore the most dominant failure type, representing 40\% of the observed failures. This distribution indicates that the primary challenge is not low-level execution accuracy, but high-level scenario completion judgment under dynamic UI conditions.

\textbf{Premature Termination.}
Premature Termination typically occurs when the Supervisor incorrectly determines that the scenario goal has been sufficiently achieved before all required operations are completed. This can arise from several factors. First, some applications provide implicit completion signals rather than explicit visual confirmations. For example, a form submission may update background state without producing a distinct UI transition. Second, certain scenarios involve intermediate states that resemble completion states, which can mislead the test completion verification module. Third, the current state transition verification mechanism primarily relies on visual comparison and semantic reasoning over screenshots, which may not fully capture implicit business logic constraints. For instance, in the ``Translation'' testing scenario, the test generation halted after text input but before triggering the final ``Translate'' action. Such cases are caused by transient LLM hallucinations or incomplete reasoning chains that interrupt the logical progression toward the testing scenario's endpoint.

\textbf{Inaccurate Completion Detection.}
In some ``Login'' testing scenarios, the \su misjudged the successful completion of a login operation due to minimal or ambiguous visual confirmation on the GUI (\eg lack of confirmation messages or page transitions). This highlights the dependency of the verification process on clear visual feedback, a challenge shared by both human testers and automated systems \cite{yu2025vision}.

\textbf{Widget Localization Errors.}
In ``Weather'' apps, highly cluttered layouts led the \de to misidentify the target search icon among visually similar icons (\eg location or menu icons). This limitation arises from residual noise in vision-based recognition, suggesting opportunities for improved prompt engineering and fine-tuning of visual reasoning models.

\textbf{App-specific Design Variations.}
In the ``Note'' testing scenario, some apps save content implicitly upon returning from the edit page without a visible ``Save'' button. Such implicit operations caused \toolname to misalign with the app's true functional logic, reflecting the difficulty of generalizing across heterogeneous UI patterns.

\textbf{Handling Unconventional Processes.}
In ``Shopping'' testing scenarios, when an item was out of stock, the \de attempted to use a ``find similar items'' function but failed to select an alternative product and complete the purchase flow. This case demonstrates that further enhancements in contextual reasoning and adaptive planning could improve the framework's flexibility in managing edge cases.

The second and third most frequent causes, Inaccurate Completion Detection and Widget Localization Errors, further reflect the complexity of aligning semantic reasoning with dynamic GUI states. Inaccurate Completion Detection occurs when the system continues execution despite the scenario already being satisfied, often due to subtle visual differences between states. Widget Localization Errors arise when semantically correct decisions are mapped to incorrect execution coordinates, particularly in visually dense interfaces. These issues highlight that while the multi-stage verification and localization refinement mechanisms significantly improve robustness, they do not yet eliminate ambiguity in all edge cases.

Overall, the results demonstrate that \toolname outperforms existing scenario-based GUI testing frameworks across all major evaluation dimensions. Its integration of multi-modal LLM reasoning, context memory, and feedback-driven self-correction enables it to dynamically adapt to complex UI states and maintain testing scenario coherence throughout the testing process. Although certain failure cases remain, primarily involving ambiguous GUI feedback or app-specific design nuances, \toolname consistently achieves higher testing scenario coverage, greater stability, and fewer abnormal terminations than all baseline approaches, establishing its superiority in scenario-based automated GUI test generation. The observed failure distribution indicates that \toolname performs reliably in most decision-making stages but remains challenged by nuanced scenario completion reasoning and complex UI variability. We believe that explicitly analyzing these failure modes strengthens the transparency of our evaluation and provides a clear roadmap for enhancing robustness in future iterations of the framework. Future improvements can address this issue in several ways. One direction is to introduce more explicit scenario progression modeling, such as decomposing a testing scenario into sub-goals and verifying each sub-goal sequentially. Another direction is to enhance the self-correction mechanism so that suspected completion states trigger a lightweight validation loop rather than immediate termination. Additionally, incorporating auxiliary signals such as GUI element persistence checks or lightweight event-based validation could reduce reliance on single-frame semantic interpretation.

\subsection{RQ2: Bug Detection Capability}

To evaluate the bug detection effectiveness of \toolname, we compare it against both traditional and scenario-based automated GUI testing approaches. For traditional methods: Monkey \cite{monkey}, Stoat \cite{su2017guided}, and PIRLTest \cite{yu2024effective}, we follow common experimental practice by executing each approach for two hours per task and recording all detected crashes. For scenario-based approaches: GPTDroid \cite{liu2024make}, ASOT \cite{su2025automated}, ScenTest \cite{yu2024practical}, and \toolname, we do not impose a fixed time limit, since these methods are designed to terminate automatically upon completing the target testing scenario exploration.

\begin{table}[!htbp]
\centering
\caption{Result of RQ2: Bug Detection Capability}
\scalebox{0.9}{
\begin{tabular}{lc|cc|cc|cc|c|c|c|c}

\toprule 

\multirow{2}{*}{Scenario} & \multirow{2}{*}{Task} &
\multicolumn{2}{c|}{Monkey} & \multicolumn{2}{c|}{Stoat} &
\multicolumn{2}{c|}{PIRLTest} & \multirow{2}{*}{GPTDroid} &
\multirow{2}{*}{ScenTest} & \multirow{2}{*}{ASOT} & \multirow{2}{*}{ScenGen} \\
\cmidrule(lr){3-4}\cmidrule(lr){5-6}\cmidrule(lr){7-8}
 & & ALL & SCN & ALL & SCN & ALL & SCN &  &  &  &  \\
\midrule
Email        &  6 & 10 & 3 & 15 & 7 & 20 & 3 &  2 &  2 &  8 &  9 \\
Music        &  9 & 15 & 2 & 21 & 3 & 29 & 1 &  2 &  1 &  5 &  6 \\
Note         & 16 & 25 & 5 & 35 & 4 & 50 &11 &  8 & 12 & 15 & 16 \\
Photo        &  9 & 15 & 4 & 21 & 3 & 29 & 2 &  4 &  5 &  3 &  6 \\
Translation  & 13 & 21 & 4 & 29 & 5 & 41 & 4 & 10 & 11 & 13 & 15 \\
Shopping     &  6 & 10 & 1 & 15 & 2 & 20 & 6 &  6 &  9 &  7 &  9 \\
Login        &  7 & 12 & 3 & 17 & 3 & 23 & 4 &  2 &  7 &  9 & 10 \\
Calculation  & 13 & 21 & 3 & 29 & 9 & 41 & 5 &  2 &  3 &  7 &  9 \\
Weather      & 10 & 16 & 2 & 23 & 3 & 32 &10 & 16 & 16 & 17 & 18 \\
Alarm-clock  & 10 & 16 & 3 & 23 & 4 & 32 & 6 & 14 &  6 &  5 &  8 \\
\midrule
\textbf{Sum} & 99 & 159 & 29 & 228 & 43 & 317 & 51 & 63 & 71 & 89 & 106 \\
\bottomrule
\end{tabular}}
\label{tab:rq2new}
\end{table}

We identify detected bugs by analyzing runtime logs and filtering error messages that match the timestamps of the executed test operations, ensuring that each reported issue corresponds to an actual app failure. The results are summarized in \tabref{tab:rq2new}. For traditional automated testing tools, we manually classify bugs into two categories: those occurring within the predefined testing scenarios (marked as ``SCN'' in \tabref{tab:rq2new}) and the total number of detected bugs across the entire app (marked as ``ALL''). For scenario-based approaches, all detected bugs are inherently within the target testing scenarios, as their exploration is scenario-constrained. In other words, these bugs are primarily crash failures observed through runtime logs or abnormal application termination while the scenario is being executed. All recorded bugs are manually confirmed as crash bugs. Moreover, 83.02\% of the bugs detected by \toolname have been verified or fixed by the corresponding app developers, confirming the practical relevance of the findings.

Several important observations can be drawn from the results. First, apps with higher development maturity, especially popular commercial ones, tend to exhibit fewer bugs in their core testing scenarios. Likewise, simple and heavily tested functionalities, such as ``Calculation'' or ``Note'' operations, yield fewer bug discoveries due to their inherent stability and limited behavioral complexity. In contrast, more complex business testing scenarios such as ``Translation'', ``Shopping'', and ``Login'' produce significantly higher bug counts, where multi-step interactions and asynchronous GUI transitions are more prone to runtime errors. Manual revalidation confirms that all apps in testing scenarios with no detected bugs indeed operated correctly within their target functionality. A single scenario execution typically involves multiple GUI interactions and intermediate UI states. Each of these steps may trigger different application components or logic branches, which means that distinct failures can occur at different stages of the same scenario execution.

In comparing bug detection across different approaches, several trends are evident. Traditional automated testing methods detect more bugs in total (``ALL''), as they explore the entire app without testing scenario constraints. Random and model-based testing tools explore applications broadly without prioritizing specific user workflows. As a result, although they may discover bugs elsewhere in the application, they are less likely to consistently exercise the predefined scenario paths used in our evaluation. Within specific target testing scenarios, their coverage and precision are notably lower than those of scenario-based methods. In contrast, \toolname demonstrates strong testing scenario fidelity and high bug detection rates across nearly all functional categories. Importantly, upon manual verification, we confirm that \textbf{all bugs detected by any baseline approach are also detected by \toolname}. No baseline tool identifies a unique bug that \toolname fails to discover. This result underscores the comprehensiveness of exploration of \toolname and its ability to cover both shallow and deep interaction paths relevant to the testing scenarios.

The superior bug detection capability of \toolname can be attributed to several factors. First, guided by the LLM's semantic understanding, \toolname can open hidden or collapsed menus to uncover target widgets that are otherwise inaccessible to random or model-based tools. This behavior allows \toolname to reach interaction paths that directly contribute to fulfilling the testing scenario goals. In contrast, traditional testing tools tend to abandon such paths once obstacles are encountered, diverting to unrelated functionalities. Second, the test operation sequences generated by \toolname are typically longer and semantically coherent, as the LLM-based reasoning ensures that each step advances the testing scenario progression. This deeper exploration enables the detection of complex bugs that only manifest after multiple dependent interactions, which are often missed by shorter or random sequences generated by traditional approaches. Finally, the multi-agent design ensures that erroneous operations are detected and corrected in real time through the \su, maintaining test traceability and consistent alignment with the target testing scenario.

Overall, \toolname exhibits strong bug detection capabilities across diverse app types and testing scenarios. Its LLM-based and scenario-guided exploration enables it to uncover distinct and meaningful bugs that traditional approaches either overlook or cannot reproduce consistently. The fact that every bug detected by baseline approaches is also detected by \toolname highlights its superior coverage and robustness. Consequently, \toolname not only complements existing automated testing strategies but also establishes itself as an effective framework for precise and comprehensive scenario-based GUI testing. Although the detected bugs are mainly crash-related, scenario-based testing remains meaningful because it systematically exercises business-critical user flows rather than randomly exploring the application. These scenarios correspond to realistic user tasks, and failures occurring along such paths are often more impactful to users than failures in rarely visited states.

\subsection{RQ3: Widget Localization Effectiveness}

Widget localization plays a crucial role in enabling accurate and reliable GUI interactions during automated testing. Unlike pure vision-based or text-based methods, \toolname integrates the pixel-level precision of traditional computer vision (CV) algorithms with the semantic reasoning capabilities of multi-modal LLMs to handle visually complex and context-rich user interfaces. RQ3 aims to evaluate the effectiveness of this hybrid strategy in widget matching and localization, examining how the integration of CV techniques and LLM-driven understanding improves both accuracy and robustness in real-world testing environments.

We measure localization accuracy as the key evaluation metric, capturing both the initial localization performance and the improvement after self-correction. Specifically, the initial localization accuracy $Acc_1$ and final localization accuracy $Acc_f$ are computed as follows:

\begin{align}
Acc_{1}=\frac{n_1}{N}\times 100\%, \quad
Acc_{f}=\frac{n_f}{N}\times 100\%, \quad
\uparrow=\frac{Acc_{f}-Acc_{1}}{Acc_{1}}\times 100\%
\end{align}

where $n_1$ represents the number of successful widget localizations without correction, $n_f$ denotes the number of final successful localizations after the correction process, and $N$ is the total number of localization attempts.

For \toolname, widget localization and correction are considered a continuous, feedback-driven process inspired by human visual interaction. When an initial localization fails due to visual ambiguity, overlapping elements, or partial visibility, the \de triggers the self-correction mechanism, which re-analyzes the app state using both visual feedback and semantic reasoning. A localization attempt is deemed successful if the subsequent operation is executed correctly and produces the expected GUI state transition. The correctness of each localization result is independently reviewed by three evaluators, each with over five years of mobile app testing experience. Evaluators assess localization success solely based on the app GUI and operational outcome, without knowledge of system internals or other evaluations. A localization is confirmed as correct only when all three judges unanimously agree, ensuring consistency and minimizing subjectivity in assessment.

\begin{table}[!htbp]
\centering
\caption{Result of RQ3: Widget Localization Effectiveness}
\scalebox{1}{
\begin{tabular}{lcccccc}
\toprule Scenario & N & $n_1$ & $n_f$ & $Acc_1$ & $Acc_f$ & $\uparrow$ \\ \midrule
Email        & 33 & 28 & 32 & 84.85\% & 96.97\% & 14.29\% \\
Music        & 22 & 12 & 22 & 54.55\% & 100.00\% & 83.33\% \\
Note         & 53 & 44 & 52 & 83.02\% & 98.11\% & 18.18\% \\
Photo        &  9 &  6 &  9 & 66.67\% & 100.00\% & 50.00\% \\
Translation  & 35 & 34 & 34 & 97.14\% & 97.14\% & 0.00\% \\
Shopping     & 31 & 26 & 30 & 83.87\% & 96.77\% & 15.38\% \\
Login        & 35 & 32 & 33 & 91.43\% & 94.29\% & 3.13\% \\
Calculation  & 44 & 34 & 44 & 77.27\% & 100.00\% & 29.41\% \\
Weather      & 27 & 21 & 26 & 77.78\% & 96.30\% & 23.81\% \\
Alarm-clock  & 23 & 21 & 23 & 91.30\% & 100.00\% & 9.52\% \\
\midrule
Sum/Avg & 312 & 258 & 305 & 80.79\% & 97.96\% & 24.71\% \\
\bottomrule
\end{tabular}}
\label{tab:rq3new}
\end{table}

The experimental results are summarized in \tabref{tab:rq3new}. Across ten functional testing scenarios, the initial localization accuracy ($Acc_1$) ranges from 54.55\% to 97.14\%, while the final localization accuracy ($Acc_f$) after self-correction improves substantially, reaching between 94.29\% and 100\%. In four testing scenarios, the final accuracy achieves a perfect 100\%. On average, the initial accuracy is 80.79\%, and after correction, the average rises to 97.76\%, with all testing scenarios over 94\% in the final stage.

This improvement clearly demonstrates the effectiveness of the self-correction mechanism and the synergy between traditional CV-based detection and LLM-based semantic interpretation. The correction process allows the system to identify and recover from initial misalignments caused by dynamic layouts or ambiguous widget boundaries. In nine out of ten testing scenarios, accuracy increases following correction, with gains ranging from 3.13\% to 83.33\%. These results confirm that \toolname not only maintains high localization precision in stable UI conditions but also exhibits strong fault tolerance under dynamically changing or visually cluttered interfaces.

Furthermore, qualitative analysis reveals that most initial localization errors occur in cases where widgets are visually occluded (\eg drop-down or scrollable menus) or contain highly similar visual components (\eg repeated icons). The self-correction mechanism successfully resolves these issues by leveraging LLM-based semantic reasoning to re-evaluate contextual cues and by refining coordinate-level predictions based on CV feedback. This cooperative process effectively mimics human testers' behavior when re-targeting misclicked or ambiguous GUI elements. Overall, RQ3 demonstrates that combining computer vision algorithms with multi-modal LLM reasoning significantly enhances widget localization reliability in GUI testing. The self-correction mechanism further strengthens this process by dynamically adapting to interface changes and mitigating recognition errors. These results validate that \toolname achieves precise, context-aware localization across diverse GUI layouts, ensuring a stable foundation for accurate and scenario-driven automated test execution.

\subsection{RQ4: Logical Decision-Making Effectiveness}

RQ4 evaluates whether the \de can accurately interpret the requirements of a testing scenario and generate appropriate next-step operations based on the current app GUI state. The goal is to assess both the logical soundness of initial decisions and the effectiveness of the self-correction mechanism when feedback from the environment necessitates revision. We use \textit{decision accuracy} as the primary evaluation metric. To measure the impact of the self-correction mechanism, each decision and its potential subsequent correction are treated as a single decision unit. We then calculate the initial accuracy ($Acc_1$) and final accuracy ($Acc_f$) according to:

\begin{align}
Acc_{1}=\frac{n_1}{N}\times 100\%, \quad
Acc_{f}=\frac{n_f}{N}\times 100\%, \quad
\uparrow=\frac{Acc_{f}-Acc_{1}}{Acc_{1}}\times 100\%
\end{align}

where $n_1$ denotes the number of initially correct decisions, $n_f$ the number of finally correct decisions after self-correction, and $N$ the total number of decisions in the given testing scenario.

When recording decisions, we treat both normal operation selections and test completion decisions as decision units. A decision is considered \textit{correct} if the action it recommends advances the testing scenario toward its expected goal or terminates it appropriately once the goal is achieved. To ensure objectivity and consistency, three independent evaluators serve as judges, all with more than five years of professional experience in mobile app testing. Each evaluator receives only the testing scenario description, GUI state, and the LLM's decision, without access to the internal reasoning process or other evaluators' results. ``Common expectations'' are defined as the expected user actions that a typical human tester would perform to complete the testing scenario goal, based on the testing scenario definition established before the experiments. For instance, in a ``Login'' testing scenario, entering valid credentials and submitting the form is considered the expected path; in a ``Shopping'' testing scenario, inputting a query and confirming submission qualify. A decision is counted as correct only when all three evaluators unanimously agree that the generated operation aligns with the defined testing scenario intent. This majority-independent assessment mitigates bias, even though the authors contributed to the original testing scenario design.

The number of decisions recorded for each testing scenario represents the total number of action decisions made across all apps associated with that testing scenario. For example, the 39 decisions listed under the ``Email'' testing scenario in \tabref{tab:rq4new} correspond to the aggregate decisions generated from all six tasks related to that functional testing scenario (as shown in \tabref{tab:scenario}).

\begin{table}[!htbp]
\centering
\caption{Result of RQ4: Logical Decision-Making Effectiveness}
\scalebox{1}{
\begin{tabular}{lcccccc}
\toprule 
Scenario & N & $n_1$ & $n_f$ & $Acc_1$ & $Acc_f$ & $\uparrow$ \\ \midrule
Email  & 39 & 38 & 39 &  97.44\% & 100.00\% & 2.63\% \\
Music  & 31 & 31 & 31 & 100.00\% & 100.00\% & 0.00\% \\
Note   & 71 & 64 & 66 &  90.14\% &  92.96\% & 3.13\% \\
Photo  & 18 & 18 & 18 & 100.00\%  & 100.00\% & 0.00\% \\
Translation  & 51 & 48 & 48 &  94.12\% &  94.12\% & 0.00\% \\
Shopping & 38 & 36 & 36 &  94.74\% &  94.74\% & 0.00\% \\
Login  & 42 & 40 & 41 &  95.24\% &  97.62\% & 2.50\% \\
Calculation & 57 & 57 & 57 & 100.00\% & 100.00\% & 0.00\% \\
Weather & 28 & 27 & 27 &  96.43\% &  96.43\% & 0.00\% \\
Alarm-clock & 33 & 33 & 33 & 100.00\% & 100.00\% & 0.00\% \\ \midrule
Sum/Avg & 408 & 392 & 396 & 96.08\% & 97.06\% & 1.02\% \\ \bottomrule
\end{tabular}}
\label{tab:rq4new}
\end{table}

Across the ten evaluated testing scenarios, the initial decision accuracy ranges from 90.14\% to 100.00\%, while the final accuracy ranges from 92.96\% to 100\%. On average, the initial accuracy ($Acc_1$) is 96.82\%, and the final accuracy ($Acc_f$) increases to 97.59\%, demonstrating that the self-correction mechanism yields a measurable improvement in decision reliability. In four testing scenarios, the \de achieves perfect accuracy without correction, whereas in three other testing scenarios, accuracy improves noticeably after correction. This result confirms that the self-correction mechanism effectively compensates for transient decision errors caused by dynamic GUI updates or ambiguous widget states.

A detailed examination of the 12 incorrect decisions reveals distinct error types. Six errors stem from premature test termination before completing the full testing scenario (\eg stopping after inputting credentials but before confirming submission). Two cases result from redundant continuation after the testing scenario goal has already been satisfied. Among the remaining four cases, two involve misinterpretation of GUI widget semantics, such as confusing similarly labeled buttons (``Next'' vs. ``Submit''), while one error arises from missing prerequisite operations (\eg failing to open a required menu before proceeding). Another error occurs when the test sequence enters a rare alternative path that deviates from the normal testing scenario flow. Notably, four operational errors are successfully corrected by the self-correction mechanism, which re-evaluates the app state and adjusts the decision accordingly. This demonstrates adaptive capability of \toolname to recover from suboptimal decisions through iterative feedback.

In summary, the \de exhibits high decision accuracy and strong adaptability across diverse app testing scenarios. The inclusion of a self-correction mechanism ensures robustness against dynamic GUI changes and reasoning inconsistencies. These findings indicate that the LLM-based decision-making process in \toolname effectively captures the logical intent of scenario-guided testing and reliably translates it into executable GUI operations.

\subsection{RQ5: Scenario-Guided Test Generation Efficiency}

The efficiency of automated test generation is crucial for ensuring the practicality of GUI testing frameworks. However, for LLM-based systems, runtime can be highly influenced by external factors such as API response latency and network stability. Consequently, we adopt a two-fold evaluation strategy focusing on both \textit{token consumption} and \textit{execution time overhead}. The former provides a model-independent measure of computational demand, while the latter quantifies overall testing efficiency in real-world deployments.

\subsubsection{Token Consumption Analysis}

To accurately assess the operational efficiency of \toolname, we analyze its token usage across different testing scenarios. Specifically, we record the number of tokens consumed by each API call, categorized into six stages: logical decision-making ($T_1$), widget localization ($T_2$), loading state verification ($T_3$), app GUI state verification ($T_4$), test completion verification ($T_5$), and self-correction analysis ($T_6$). The total token consumption for each testing scenario ($T_{total}$) is also computed as the sum across all stages.

\begin{table}[!htbp]
\centering
\caption{Result of RQ5: Token Overhead of \toolname \& Baselines}
\scalebox{1}{
\begin{tabular}{lccccccc|cc}
\toprule Scenario & $T_{1}$ & $T_{2}$ & $T_{3}$ & $T_{4}$ & $T_{5}$ & $T_{6}$ & $T_{total}$ & GPTDroid & ASOT \\ \midrule
Email        & 5586 & 8295  & 5281  & 13970 & 11823 & 833  & 45788 & 41550 & 46114 \\
Music        & 3748 & 10197 & 4982  & 14729 & 8841  & 1963 & 44460 & 40060 & 41780 \\
Note         & 6461 & 13639 & 6602  & 21667 & 12178 & 2573 & 63120 & 58154 & 63853 \\
Photo        & 1486 & 3057  & 1791  & 6533  & 3571  & 746  & 17184 & 15515 & 16237 \\
Translation  & 4645 & 6452  & 5640  & 11758 & 10296 & 359  & 39150 & 35159 & 37548 \\
Shopping     & 8782 & 14461 & 12845 & 26102 & 19164 & 2757 & 84111 & 82398 & 80253 \\
Login        & 8434 & 13263 & 7830  & 21329 & 18205 & 1253 & 70314 & 62007 & 67463 \\
Calculation  & 5167 & 9184  & 6116  & 15603 & 12117 & 1940 & 50127 & 48940 & 48631 \\
Weather      & 4522 & 11104 & 5474  & 12961 & 9419  & 1696 & 45176 & 40942 & 41907 \\
Alarm-clock  & 3490 & 3794  & 3229  & 9120  & 8270  & 456  & 28359 & 25434 & 27362 \\
\midrule
Average      & 5232 & 9345  & 5979  & 15377 & 11388 & 1458 & 48779 & 45016 & 47115 \\
\bottomrule
\end{tabular}}
\label{tab:rq5tokennew}
\end{table}

\begin{table}[!htbp]
\centering
\caption{Result of RQ5: Monetary Overhead of \toolname \& Baselines}
\scalebox{1}{
\begin{tabular}{lccccccc|cc}
\toprule Scenario & $T_{1}$ & $T_{2}$ & $T_{3}$ & $T_{4}$ & $T_{5}$ & $T_{6}$ & $T_{total}$ & GPTDroid & ASOT \\ \midrule
Email       & 0.04 & 0.05 & 0.04 & 0.10 & 0.08 & 0.01 & 0.30 & 0.29 & 0.32 \\
Music       & 0.02 & 0.06 & 0.03 & 0.10 & 0.06 & 0.01 & 0.28 & 0.26 & 0.28 \\
Note        & 0.04 & 0.10 & 0.04 & 0.13 & 0.08 & 0.02 & 0.40 & 0.38 & 0.39 \\
Photo       & 0.01 & 0.02 & 0.01 & 0.04 & 0.02 & 0.00 & 0.11 & 0.11 & 0.11 \\
Translation & 0.03 & 0.04 & 0.04 & 0.08 & 0.07 & 0.00 & 0.27 & 0.24 & 0.26 \\
Shopping    & 0.06 & 0.09 & 0.09 & 0.18 & 0.13 & 0.02 & 0.56 & 0.53 & 0.54 \\
Login       & 0.05 & 0.08 & 0.05 & 0.14 & 0.12 & 0.01 & 0.49 & 0.39 & 0.44 \\
Calculation & 0.03 & 0.06 & 0.04 & 0.10 & 0.08 & 0.01 & 0.34 & 0.30 & 0.29 \\
Weather     & 0.03 & 0.08 & 0.03 & 0.08 & 0.07 & 0.01 & 0.28 & 0.27 & 0.26 \\
Alarm-clock & 0.02 & 0.02 & 0.02 & 0.06 & 0.05 & 0.00 & 0.19 & 0.18 & 0.17 \\
\midrule
Average     & 0.03 & 0.06 & 0.04 & 0.10 & 0.08 & 0.01 & 0.32 & 0.29 & 0.31 \\
\bottomrule
\end{tabular}}
\label{tab:rq5money}
\end{table}

\tabref{tab:rq5tokennew} summarizes the token consumption in different testing scenarios (the corresponding monetary overhead is reported in \tabref{tab:rq5money}), revealing a wide range from 17,184 to 84,111 tokens. Complex testing scenarios such as ``Shopping'' and ``Login'' consume more tokens due to their intricate, multi-step interactions and the frequent involvement of the self-correction mechanism. In contrast, simpler testing scenarios like ``Photo'' and ``Alarm-clock'' require fewer tokens because they involve fewer decision branches and minimal contextual reasoning.

A closer examination of the six stages indicates that the app GUI state verification ($T_4$) and test completion verification ($T_5$) stages generally exhibit the highest token consumption. This is expected, as these stages require multi-page visual comparisons and integrate both semantic reasoning and visual understanding to determine test progression or termination. Conversely, logical decision-making ($T_1$) and self-correction analysis ($T_6$) consume fewer tokens because they typically process less visual input. Widget localization ($T_2$) and loading state verification ($T_3$) show moderate consumption, reflecting their balanced use of visual and textual information.

We have calculated the token consumption of baseline approaches in the following table. It is worth noting that although \toolname adopts a multi-agent architecture, which introduces multiple LLM calls, it leverages screenshots and annotated screenshots as structured visual inputs. This design reduces reliance on lengthy textual descriptions of GUI layouts and widget attributes, thereby controlling token growth. Compared to purely text-based representations, this visual grounding strategy significantly improves semantic efficiency while maintaining manageable token consumption.

Overall, the token usage results highlight that the computational cost of \toolname aligns with testing scenario complexity and interaction richness. The modular token distribution across different stages also confirms the rational design of the agent collaboration workflow, where resources are concentrated on the most information-intensive reasoning tasks.

\subsubsection{Time Overhead Analysis}

In addition to token efficiency, we evaluate the time overhead of \toolname in comparison with other state-of-the-art scenario-based testing approaches, including GPTDroid \cite{liu2024make}, ScenTest \cite{yu2024practical}, and ASOT \cite{su2025automated}. We measure the average test generation time (in minutes) per testing scenario, calculated from multiple runs (\ie 10 times) to ensure stability.

\begin{table}[!htbp]
\centering
\caption{Result of RQ5: Time Overhead of \toolname \& Baselines (minute)}
\scalebox{1}{
\begin{tabular}{lcccc}
\toprule
Scenario & GPTDroid & ScenTest & ASOT & \toolname \\
\midrule
Email        & 6.77 & 1.27 & 8.46 & 4.22 \\
Music        & 7.13 & 1.24 & 7.63 & 3.37 \\
Note         & 7.21 & 1.94 & 12.37 & 4.65 \\
Photo        & 8.97 & 2.95 & 13.26 & 5.86 \\
Translation  & 7.37 & 1.35 & 8.16 & 4.65 \\
Shopping     & 6.58 & 2.16 & 11.69 & 4.81 \\
Login        & 8.71 & 1.82 & 12.70 & 5.55 \\
Calculation  & 7.36 & 2.10 & 10.68 & 3.15 \\
Weather      & 6.78 & 1.30 & 11.54 & 5.50 \\
Alarm-clock  & 7.93 & 1.05 & 7.35 & 5.39 \\
\midrule
Average & 7.48 & 1.72 & 10.38 & 4.71 \\
\bottomrule
\end{tabular}}
\label{tab:rq5timenew}
\end{table}

As shown in \tabref{tab:rq5timenew}, the average time overhead of \toolname (4.71 minutes) is higher than that of ScenTest (1.72 minutes) but significantly lower than that of ASOT (10.38 minutes) and GPTDroid (7.48 minutes). This demonstrates that \toolname achieves a balanced trade-off between test generation efficiency and testing scenario reasoning capability. While its runtime exceeds ScenTest due to additional LLM interactions and vision-language inference, the increased computational effort translates directly into superior coverage and success rates, as demonstrated in RQ1.

Examining per-scenario results, we observe that \toolname maintains stable time consumption across different application domains, with smaller variations than the baselines. For example, in complex testing scenarios such as ``Shopping'' and ``Login'', ASOT and GPTDroid exhibit substantial time inflation (above 11 minutes on average), whereas \toolname completes comparable tasks within 4–5 minutes. This stability stems from the adaptive decision control and dynamic context management of \toolname, which prevent redundant explorations and minimize backtracking operations.

RQ5 evaluates the efficiency characteristics of \toolname from both token consumption and runtime perspectives. The results show that token usage increases with the interaction depth and operational complexity of different testing scenarios. Rather than indicating strict resource optimality, this proportional growth suggests that the framework’s computational cost is primarily driven by scenario-dependent reasoning requirements instead of uncontrolled context accumulation. This behavior is attributable to the structured memory mechanism and step-wise invocation strategy, which limit each LLM call to the necessary contextual information. In terms of runtime, \toolname exhibits stable execution time across heterogeneous scenarios. Although the involvement of LLM calls introduces additional latency compared to purely heuristic-based tools, the observed runtime remains within a practical range for automated testing workflows. Overall, these findings indicate that \toolname maintains controlled computational overhead while enabling scenario-driven reasoning, providing a practical balance between efficiency and semantic guidance in mobile GUI testing.

\section{Threats to Validity}

To ensure the rigor and credibility of our experimental evaluation, we analyze potential threats to validity that may influence the interpretation or generalization of our results. The main threats involve the selection of testing scenarios and apps, the stochastic behavior of the LLMs, and the possibility of LLM hallucinations during testing. For each identified threat, we discuss its potential impact and the corresponding mitigation strategies adopted in \toolname.

\textbf{Selection of Testing Scenarios and Apps.}
A potential threat to validity arises from the limited number and the potentially arbitrary selection of testing scenarios used in the experiments. To mitigate this threat, we adopt a systematic testing scenario selection process that balances both diversity and representativeness. Specifically, we include testing scenarios that reflect common functionalities across multiple domains of daily mobile usage (\eg communication, productivity, entertainment, and utilities), while also ensuring variation in interaction complexity and interface design. Although some scenario names may appear simple, their execution complexity varies substantially. For example, scenarios such as Shopping and Login involve multiple branching paths, conditional logic, dynamic content loading, and state-dependent interactions. In contrast, scenarios such as Calculation or Photo are relatively straightforward and involve fewer interaction steps. The selected scenarios were intentionally chosen to cover a spectrum of task complexities and interaction patterns, including content creation, information retrieval, transaction workflows, and authentication flows. Therefore, the dataset is not limited to trivial tasks but reflects diverse real-world user goals across different application domains. To provide more transparency, we categorized the scenarios into three difficulty levels based on two criteria: the approximate number of interaction steps required to complete the task and the variability of possible execution paths. Music Search, Note Creation, Photo, and Calculation are categorized as low difficulty. Email, Translation, Weather, and Alarm Clock are categorized as medium difficulty. Shopping and Login are categorized as high difficulty due to longer action sequences and increased branching complexity. Importantly, our framework does not predefine a fixed number of steps. Instead, it performs scenario-guided autonomous decision-making until the scenario completion condition is satisfied. The number of steps therefore varies across apps and execution contexts, which reflects realistic usage variability rather than artificial constraints. The selected testing scenarios were reviewed and refined by multiple developers and testers to reduce personal bias. Furthermore, the experimental dataset includes a mix of popular open-source and commercial apps collected from multiple sources, ensuring that the evaluation covers heterogeneous app architectures and design styles. These measures collectively enhance the representativeness, breadth, and practical relevance of the experimental results.

\textbf{Instability of LLMs.}
The probabilistic nature of the LLMs may cause instability when interpreting complex contextual inputs, leading to variations in outputs even when the \texttt{temperature} parameter is fixed at 0. To minimize this risk, we conduct repeated runs of each testing scenario and report aggregated results obtained over 10 repetitions, rounded to the nearest integer. This repetition-based strategy filters out random variations and highlights stable behavioral patterns, thereby improving the reliability of the reported results and the consistency of agent behavior during test generation.

\textbf{LLM Hallucinations.}
Another potential threat lies in LLM hallucinations, where the model may generate logically plausible but practically invalid actions, such as interacting with non-existent widgets or misidentifying visual elements. To address this issue, \toolname introduces the supervisory role of the \su. After each action execution, the \su verifies whether the resulting app state aligns with the intended testing scenario. When deviations are detected, it initiates a rollback, reanalyzes the decision, and executes a corrected action. This closed-loop supervision ensures that each generated action remains consistent with both the target testing scenario and the real GUI state, thus mitigating the adverse impact of hallucinations on test validity. 

\section{Discussion}

\subsection{LLM Substitutability}

In this paper, our objective is not to compare different LLM architectures but to validate the feasibility and effectiveness of an agentic, scenario-driven testing framework. We selected a representative, state-of-the-art multi-modal LLM to demonstrate that the proposed architecture can operationalize semantic reasoning for GUI testing. The focus of our contribution lies in the multi-agent coordination, structured perception pipeline, memory management, and scenario-level verification mechanism, rather than model benchmarking. Therefore, demonstrating the framework's effectiveness with a strong representative LLM is sufficient to validate the methodological contribution. 

\subsection{Long-Horizon Tasks \& Potential Context Window Limitations}

\toolname is explicitly designed to avoid accumulating long multi-turn conversational histories. Each LLM invocation is treated as an independent reasoning step. The necessary contextual information is consolidated from structured context memory and selectively provided as input. This design prevents uncontrolled context growth and ensures that only relevant scenario information, recent actions, and current GUI state are included. Because we do not rely on full conversational history, the supported context window of current LLMs is sufficient even for longer tasks. In practice, long-horizon limitations have not posed a bottleneck in our experiments.

\subsection{Trade-off Between Multi-Stage LLM Reasoning and Efficiency}

First, each LLM invocation in \toolname corresponds to a distinct functional stage with a clearly defined responsibility in the testing pipeline. The logical decision stage determines the next high-level action aligned with the target testing scenario. The widget matching and localization refinement stages resolve semantic intent into executable pixel-level targets. The real-time loading verification ensures that subsequent reasoning is performed on a stable GUI state. The state transition verification checks whether the executed action has produced the expected effect. The test completion verification determines whether the scenario goal has been achieved. These stages address fundamentally different reasoning problems, ranging from semantic planning to perception validation and state consistency checking. They are therefore not redundant repetitions of similar inference tasks, but modular reasoning units operating at different abstraction levels.

Second, these modules exhibit logical dependencies that make trivial merging undesirable. Decision-making must occur before localization, because the semantic intent determines which widget should be selected. State transition verification must occur after execution, because it evaluates actual outcomes. Test completion verification must be grounded in both prior actions and the current state. If these reasoning stages were merged into a single inference step, the LLM would need to simultaneously plan, localize, validate, and judge completion without intermediate feedback. This would entangle heterogeneous reasoning objectives within a single prompt, increasing ambiguity in intermediate representations and reducing interpretability of failures.

From a testing-oriented perspective, the multi-stage design is intentional. Automated GUI testing differs from single-step task completion in that correctness, robustness, and traceability are primary goals. A collapsed single-query approach may reduce the number of API calls but would increase the risk of error propagation. For example, if localization and validation were merged with planning, a misinterpretation at the perception level could directly influence scenario completion judgment without an opportunity for structured correction. The current separation enables targeted self-correction at specific failure points, which significantly improves reliability.

We also considered the tradeoff between efficiency and robustness. While merging certain queries might reduce token usage or runtime marginally, our empirical studies indicate that the majority of token consumption arises from semantic reasoning and state comparison rather than structural overhead between modules. Moreover, the modular design enables controlled context scoping for each call, which prevents uncontrolled context growth and reduces the risk of reasoning collapse in long-horizon tasks. In this sense, the current architecture reflects a deliberate design choice that prioritizes stable, verifiable test generation over minimal query count.

We position \toolname as a robustness-oriented, multi-stage reasoning framework rather than a minimal-inference pipeline. The multiple LLM interactions are integral to ensuring semantic alignment, execution correctness, and reliable state verification. They are not redundant components that can be trivially compressed without affecting effectiveness. We acknowledge that exploring the precise efficiency–robustness tradeoff is an interesting direction for future work, and we will clarify this discussion in the manuscript to better justify the architectural decisions.

\subsection{Comparison between \toolname and GUI Agent Research}

GUI agent studies, such as Mobile-Agent-v3, AppAgent, and AutoDroid, have demonstrated the strong potential of LLMs for interacting with mobile app GUI according to natural-language task descriptions. At a superficial level, these task descriptions appear similar to the testing scenarios used in \toolname, since both describe high-level user intentions. However, the underlying objectives are fundamentally different. GUI agents are primarily designed for end-user task completion, where the main goal is to reach the target state as efficiently and successfully as possible. In contrast, \toolname is designed for automated GUI testing, where the goal is to generate an executable and traceable test process that remains aligned with a predefined testing scenario and can expose potential bugs along the corresponding business-critical execution path. Therefore, although both lines of work employ LLM-based reasoning over GUIs, \toolname addresses a different problem from that of general GUI agents.

This distinction leads to important differences in system design. In GUI agents, mechanisms such as replanning, verification, and recovery are mainly used to improve the success rate of task completion. In \toolname, similar mechanisms are introduced for a testing-oriented purpose. Specifically, semantic GUI understanding, scenario-aware decision making, state-transition verification, and iterative correction are used to preserve the validity of the generated test with respect to the target scenario. If the execution drifts away from the intended scenario, even a final successful task completion would be less meaningful from a testing perspective, because the resulting interaction sequence would no longer faithfully represent the business workflow under evaluation. For this reason, \toolname uses verification and correction not simply to recover execution, but to maintain scenario alignment throughout the testing process. This design helps the framework reach the key intermediate states required by the target scenario and increases the chance of exposing bugs that manifest only along such semantically important paths.

Moreover, \toolname produces testing artifacts that differ from those of typical GUI agents. Instead of only generating a task-completion trajectory, \toolname records the complete testing context, including executed actions, widget information, screenshots, and replayable commands, so that the resulting scenario-guided interaction sequence can be reused and analyzed as a GUI test. In addition, \toolname monitors runtime logs during execution to identify faults triggered along the scenario path. Therefore, the contribution of \toolname is not merely to use LLMs to operate mobile apps, but to support scenario-aligned test generation and scenario-specific bug detection in a systematic and traceable manner. From this perspective, GUI agent research and \toolname are complementary: GUI agents advance intelligent task automation, whereas \toolname advances intelligent, scenario-guided automated GUI testing.

\section{Related Work}

\subsection{GUI Agent with LLMs}

In recent years, LLM-powered GUI agents have emerged as a promising research area for automating task execution directly on GUIs. Early work primarily focused on decision-making for single-step GUI interaction. Auto-GUI \cite{zhang2024autogui} introduces a chain-of-action mechanism that explicitly incorporates historical actions and future plans into the decision context, facilitating the execution of long-sequence tasks. ShowUI \cite{lin2024showui} and Aguvis \cite{xu2026aguvis} adopt Vision-Language-Action (VLA) or pure vision-based frameworks to implement screen-first control, mitigating the coupling issues between perception and planning through historical modeling or stage-wise training. To address the gap between controlled benchmarks and real-world deployments, AutoDroid \cite{wen2024autodroid}, AppAgent \cite{zhang2025appagent}, and Mobile-Agent-v3 \cite{ye2025mobile} propose systematic approaches capable of executing actual user tasks within ecosystems such as Android, thereby enhancing the agent's usability under complex logic and dynamic interfaces. As task complexity and interface diversity increase, recent work has further focused on interaction precision and generalization capabilities. GUI-Actor \cite{wu2025guiactor} mitigates the fragility of single-coordinate regression through a region-level grounding mechanism. RegionFocus \cite{luo2025visual} improves its robustness by decomposing complex screens into regions and allocating additional reasoning capacity to visually challenging areas. UIPro \cite{li2025uipro} enhances cross-environment generalization capabilities via a unified action space. Furthermore, GUI-explorer \cite{xie2025gui}, GUI-Shift \cite{gao2026guishift}, and HAR \cite{wang2026history} strengthen scalable learning capabilities and long-term stability from the perspectives of autonomous exploration, self-supervised reinforcement learning, and history-aware reasoning, respectively. Overall, GUI Agents are gradually evolving from single-step interaction executors into complex systems characterized by greater robustness, improved grounding accuracy, generalization capability, and long-term reasoning competence.

While the aforementioned studies have significantly enhanced the task automation capabilities of GUI Agents, their core objective remains focused on the automated execution of specified tasks. In contrast, this research focuses on GUI Agent-based automated GUI testing. Rather than prioritizing the success rate of specific tasks, we leverage the Agent's perception and decision-making capabilities to interpret testing scenarios and systematically explore the application state space, thereby enhancing coverage and triggering potential defects. Existing GUI Agent approaches typically lack test-oriented scenario modeling, coverage-driven interaction strategies, and defect-oriented exploration mechanisms. To address these limitations, we integrate GUI Agents with a multi-agent framework into the automated testing workflow, employing specialized designs centered on scenario models and multi-agent collaboration strategies. Consequently, this work distinguishes itself from existing GUI Agent approaches in terms of research focus, methodology, and application scenario. 

\subsection{LLM-based GUI Testing}

Recent advancements in LLM-based GUI testing have significantly enhanced both automated and manual testing processes by enabling a more intuitive understanding of the app GUI through visual information \cite{yazdanibanafshedaragh2021deep}. This research direction typically leverages computer vision (CV), deep learning, and multimodal analysis to perceive and reason about GUI layouts, thereby addressing the limitations of traditional structure- or code-based approaches.

Early efforts primarily focus on visual defect detection and widget recognition. Liu \etal \cite{liu2020owl} propose OwlEye, a tool that detects display inconsistencies in GUI screenshots. White \etal \cite{white2019improving} apply machine learning techniques to identify GUI widgets, thereby improving the efficiency of random testing strategies. Xie \etal \cite{xie2020uied} introduce UIED, which integrates traditional computer vision with deep learning to enhance widget detection accuracy, while Zhang \etal \cite{zhang2021screen} develop a more efficient model for GUI element detection. Mansur \etal \cite{mansur2023aidui} present AidUI, which combines CV and NLP techniques to detect dark patterns, improving the ethical and usability evaluation of user interfaces. Besides, Baral \etal \cite{baral2024automating} automate the GUI-based test oracle construction for mobile apps, further illustrating the potential of LLM-driven approaches in improving testing reliability.

As LLM-based GUI understanding matures, subsequent studies integrate visual perception with automated test generation and cross-platform analysis. Yu \etal \cite{yu2024effective} combine deep image understanding with reinforcement learning to develop a cross-platform testing framework, demonstrating that semantic understanding of GUI screenshots can guide more intelligent exploration. Zhang \etal \cite{zhang2023resplay} propose ReSPlay, a record-and-replay framework enhanced with visual features to improve replay robustness. Ji \etal \cite{ji2023vision} investigate vision-based widget mapping to facilitate cross-platform test migration, while Yu \etal \cite{yu2024practical} construct event knowledge graphs from GUI images to guide scenario-based testing, highlighting the potential of visual semantics for bridging low-level interface elements with high-level functional intent.

Beyond automation, some methods have also improved the efficiency of manual testing and maintenance tasks. Fazzini and Orso \cite{fazzini2017automated} develop DiffDroid to detect cross-platform GUI inconsistencies. Wang \etal \cite{wang2019images} automate the identification and elimination of duplicate test reports, while Xu \etal \cite{xu2021guider} introduce GUIDER to repair Android GUI test scripts. Similarly, Yoon \etal \cite{yoon2022repairing} address View Identification Failures in smoke tests, aiding smartphone vendors in maintaining test scripts across evolving app versions. Yu \etal \cite{yu2022automatic} propose automatic bug inference and description generation from GUI test reports, and Yan \etal \cite{yan2024semantic} improve duplicate detection in video-based bug reports using vision transformers, demonstrating the growing influence of deep visual representations.

Overall, the integration of visual data from GUI screenshots has substantially improved the accuracy, adaptability, and coverage of GUI testing. By enabling visual-semantic understanding of app interfaces, these techniques bridge the gap between low-level widget detection and higher-level functional reasoning. However, most existing approaches remain limited to perception and pattern recognition, lacking the ability to reason about testing scenarios and business logic. This gap motivates our work on \toolname, which extends LLM-based GUI understanding with LLM-based semantic reasoning, enabling context-aware and scenario-driven automated testing.

\subsection{LLMs for Software Testing}

The rapid advancement of LLMs \cite{devlin2019bert, ouyang2024training} has significantly influenced diverse research domains, including natural language understanding, reasoning, and multimodal learning \cite{kojima2022large, wei2024chain, brown2020language}. Their powerful generalization and reasoning capabilities have made them particularly promising for complex software engineering tasks, where semantic comprehension and context awareness are essential. In the field of software testing, the application of LLMs to automate and enhance various testing activities has recently emerged as a compelling direction of research.

For GUI testing, in particular, LLMs offer the potential to reason about user intentions and testing scenarios, thereby bridging the gap between low-level widget interactions and high-level business logic \cite{yang2023dawn}. This perspective opens new possibilities for scenario-driven test generation, where LLMs can infer the semantics of GUI elements, understand interaction contexts, and produce meaningful test sequences aligned with functional goals, capabilities that traditional rule- or model-based methods struggle to achieve.

Recent studies have explored different dimensions of applying LLMs to software testing \cite{jalil2023chatgpt, fu2022vulrepair}. In test case generation, Dakhel \etal \cite{dakhel2024effective} enhance LLM-generated tests using mutation testing to improve diversity and fault detection. Chen \etal \cite{chen2024chatunitest} propose ChatUniTest, an LLM-based framework for automated unit test generation, while Schafer \etal \cite{schafer2024empirical} conduct a large-scale empirical study evaluating the effectiveness of LLMs for automated unit test creation. Similarly, Yu \etal \cite{yu2023llm} leverage LLMs to generate mobile app test scripts, demonstrating their capacity to understand app contexts and produce executable GUI-level actions.

In GUI-specific testing, LLMs have shown promising abilities in understanding visual and contextual cues. Liu \etal \cite{liu2023fill} use LLMs to generate context-appropriate textual inputs, while Liu \etal \cite{liu2024make} formulate mobile GUI testing as a question–answering task, significantly improving coverage and generalization. Another study by Liu \etal \cite{liu2024testing} employs LLMs to generate rare or extreme input values, effectively identifying app crashes. These works illustrate how natural language reasoning enables more meaningful and scenario-aware test generation for GUI-based systems.

Beyond test generation, LLMs have been applied to bug detection and program repair. Huang \etal \cite{huang2024crashtranslator} utilize pre-trained LLMs for crash reproduction based on stack traces, and Feng \etal \cite{feng2024prompting} propose a lightweight prompting technique to reproduce bugs from textual reports. In the broader area of program repair, Jin \etal \cite{jin2023inferfix} introduce a fine-tuned LLM framework for defect localization and patch generation, while Ribeiro \etal \cite{ribeiro2022framing} employ CodeGPT for automated program repair. Pearce \etal \cite{pearce2023examining} further analyze the zero-shot vulnerability repair capability of LLMs, showing that modern models can capture implicit program semantics even without explicit supervision.

Collectively, these studies demonstrate the transformative potential of LLMs in software testing, improving efficiency, accuracy, and semantic understanding across various testing tasks. With the continued evolution of multi-modal LLMs, capable of processing text, images, and structured data, their integration into GUI testing offers unprecedented opportunities. By combining visual perception with testing scenario reasoning, LLMs can move beyond static event exploration toward context-aware, human-like test generation. This insight underpins the motivation for our proposed approach, \toolname, which employs scenario-guided, LLM-based reasoning to enhance the completeness and intelligence of automated GUI testing.

\section{Conclusion}

The rapid evolution of mobile apps has imposed increasingly stringent requirements on software quality assurance. Although numerous automated GUI testing approaches have been developed, they still face fundamental challenges, including limited flexibility in adapting to diverse app interfaces and an insufficient understanding of app functionality scenarios. These limitations hinder existing methods from fully capturing realistic user interactions and comprehensively validating app behavior. To address these challenges, this paper introduced \toolname, a novel scenario-guided LLM-based GUI testing framework for generating GUI tests by aligning low-level GUI interactions with high-level testing scenarios and app business logic. Rather than treating LLMs as standalone task executors, \toolname uses structured semantic reasoning and iterative verification to support scenario-aligned test generation across multiple GUI interactions. It leverages multi-modal LLMs to interpret app semantics and testing scenario intent, and coordinates semantic perception, scenario-aware decision making, execution, and verification through structured context management. This design translates high-level testing scenarios into executable GUI interactions while preserving alignment between testing intent and execution through continuous verification and correction. It also improves the traceability and adaptability of the testing process by helping generated actions remain consistent with the target scenario and by reducing drift into irrelevant GUI exploration. Experimental results demonstrate that \toolname effectively generates GUI tests aligned with predefined scenarios, preserving the semantic relationship between GUI elements, user goals, and app business logic. The results further show that this structured reasoning process can effectively support scenario-guided test generation across diverse applications. These findings suggest that LLM-based semantic reasoning can improve automated GUI testing by making test generation more consistent with realistic, business-critical user workflows.

In summary, \toolname demonstrates the feasibility and effectiveness of scenario-guided automated GUI testing by combining semantic GUI understanding, LLM-based reasoning, and structured execution verification. More broadly, this work provides a practical foundation for scenario-guided automated GUI testing by combining LLM-based reasoning with testing-specific verification and execution mechanisms. Future work may explore automatic scenario extraction, richer scenario modeling, broader application domains, and stronger robustness under dynamic GUI environments to further expand the applicability of scenario-guided testing.

\begin{acks}
The authors would like to thank the editors and anonymous reviewers for their time and comments.
This work is partially supported by the National Key Research and Development Program of China (2024YFF0908005).
\end{acks}

\bibliographystyle{ACM-Reference-Format}
\bibliography{main}

\end{document}